\documentclass[aps,prx, twocolumn,superscriptaddress,nopacs,amsmath,amssymb,final,letterpaper]{revtex4}

\usepackage[utf8]{inputenc}
\usepackage{calc}
\usepackage{graphicx}
\usepackage{amsmath,amssymb,amsthm}
\usepackage{txfonts}
\usepackage{bm}
\usepackage{color}
\usepackage{hyperref}
\usepackage{algcompatible}
\usepackage{multirow}

\usepackage{pdfpages}
\graphicspath{{./figs/}} 

\begin{document}
\author{Laurent H\'{e}bert-Dufresne}
\author{Joshua A. Grochow}
\affiliation{Santa Fe Institute, Santa Fe, NM 87501, USA}
\author{Antoine Allard}
\affiliation{Departament de F\'isica Fonamental, Universitat de Barcelona, Barcelona 08028, Spain}

\title{Multi-scale structure and topological anomaly detection via a new network statistic:\\ The onion decomposition}

\begin{abstract}
We introduce a new network statistic that measures diverse structural properties at the micro-, meso-, and macroscopic scales, while still being easy to compute and easy to interpret at a glance. Our statistic, \emph{the onion spectrum}, is based on the \emph{onion decomposition}, which refines the $k$-core decomposition, a standard network fingerprinting method. The onion spectrum is exactly as easy to compute as the $k$-cores: It is based on the stages at which each vertex gets removed from a graph in the standard algorithm for computing the $k$-cores. But the onion spectrum reveals much more information about a network, and at multiple scales; for example, it can be used to quantify node heterogeneity, degree correlations, centrality, and tree- or lattice-likeness of the whole network as well as of each $k$-core. Furthermore, unlike the $k$-core decomposition, the combined degree-onion spectrum immediately gives a clear local picture of the network around each node which allows the detection of interesting subgraphs whose topological structure differs from the global network organization. This local description can also be leveraged to easily generate samples from the ensemble of networks with a given joint degree-onion distribution. We demonstrate the utility of the onion spectrum for understanding both static and dynamic properties on several standard graph models and on many real-world networks.
\end{abstract}

\maketitle

\section{Introduction}

The ever-growing literature on complex networks is a testament both to the ubiquity of networks as conceptual tools, and to the lack of a definitive toolbox. The recent explosion in available data, both in the variety of sources (e.g., social networks, biological networks, infrastructure or technological systems) and in the sheer size of the datasets, stresses the need for good methods to analyze and synthesize information about the structure of networks.

Although there is already a plethora of metrics and methods to study networks \cite{Newman2010}, there is still a need for multi-scale metrics. For example, degree distribution \cite{Barabasi1999} and local clustering \cite{Watts1998} are incredibly simple and informative but only capture microscopic, local features of networks. Modularity (e.g., \cite{Newman2006,Newman2012}) and other community structure properties probe the meso-scale organization of networks but are still ill-defined, as evidenced by the lack of a common definition for network communities \cite{Fortunato2010}. Macroscopic measures such as betweenness centrality \cite{Goh2001}, eigenvector centrality \cite{Newman2010}, and mean shortest path length \cite{Watts1998} can characterize the role of a given node in the overall network structure. However they demand significant computational effort and rarely map back to a good local understanding of how the network might be constructed. There is a dire need for tools that are easy to compute and that complement existing tools by characterizing networks at multiple scales at a glance.

As it turns out, the answer was perhaps hidden all along in a popular algorithm: The $k$-core decomposition \cite{Seidman,Batagelj}. We generalize this algorithm both to characterize networks as a whole and to detect some of their interesting, unique subgraphs. We will discuss how our new method can be described in terms of local rules, yielding the most structurally constraining network connection process to date. Finally, we show how these constrained networks can be used to better capture a dynamical process than existing random network models.

\subsection{The Onion Decomposition}
The $k$-core decomposition is a network pruning method whose goal is to separate a network into a succession of nested cores, effectively defining a center and periphery to the network. The method is based on the concept of $k$-cores: The maximal sub-network within which every node has degree \textit{at least} $k$. Nodes that are in the $k$-core but not in the $(k+1)$-core can be said to belong to the $k$-shell and are thus of \textit{coreness} $k$. Lower $k$-shells can be thought of as being less central; the network is then viewed as a series of progressively denser and more central cores.

A useful metaphor relates the $k$-core decomposition to the peeling of an onion \cite{AlvarezHamelin}: One first removes all nodes of degree 1, then nodes who are now of degree one following the removal of the first onion skin, and so on until all nodes left are of degree greater than one and thus compose the $2$-core. The peeling then starts again, now removing nodes of degree at most 2. While the mathematical definition of the $k$-core is elegant, we argue that much more information lies in the decomposition process than in the final results. In other words, we aim to study how the speed at which one can peel the network into cores is related to its structure. 

We thus introduce the concept of \textit{layers}: How many peeling passes are needed to reach a given node. For instance, nodes of the $k$-shell belong to its first local layer if they are of degree exactly $k$ within the $k$-core, or to its second local layer if they are of degree at most $k$ only after the removal of the first layer. %We can then define the global layer of a node as its layer regardless of the shell to which it belongs.

\begin{figure*}[ht!]
    \centering
    \includegraphics[trim=0.0cm 4.0cm 0.0cm 0.0cm, clip=true, width=0.3\linewidth]{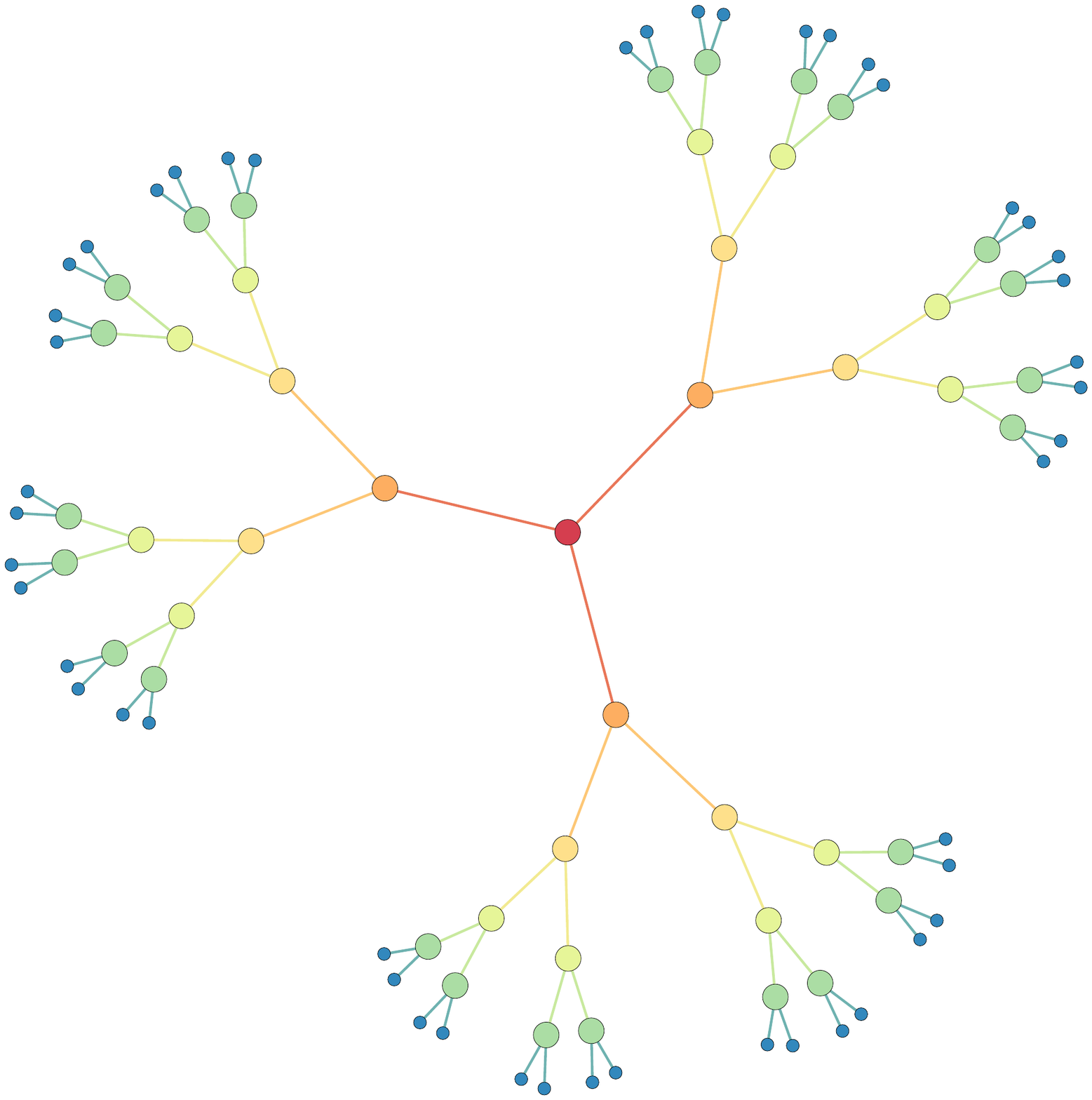} 
    \includegraphics[trim=0.0cm 0.0cm 0.0cm 0.0cm, clip=true, width=0.3\linewidth]{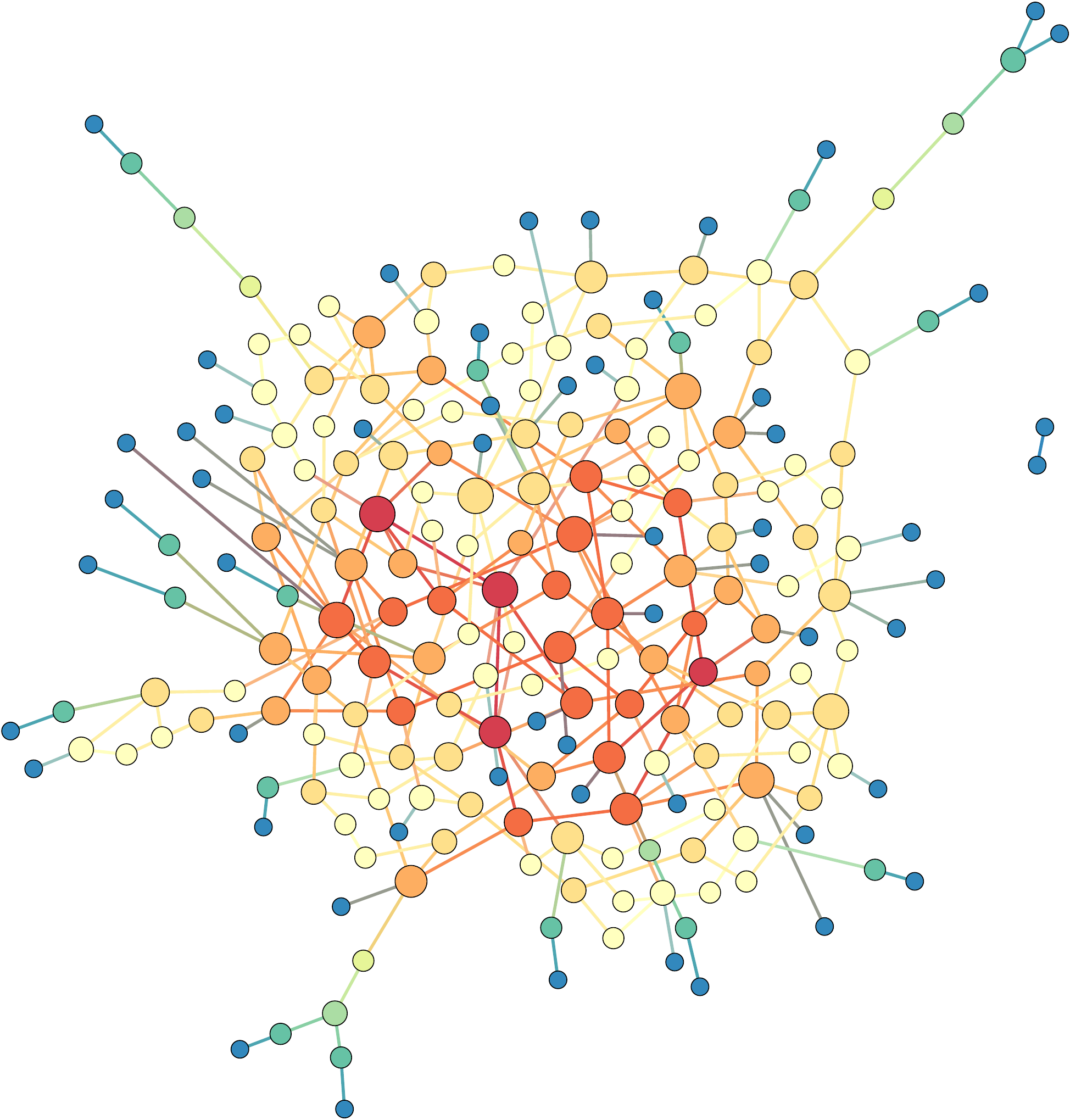} 
    \includegraphics[trim=-1.0cm 4.0cm 0.0cm 0.0cm, clip=true, width=0.3\linewidth]{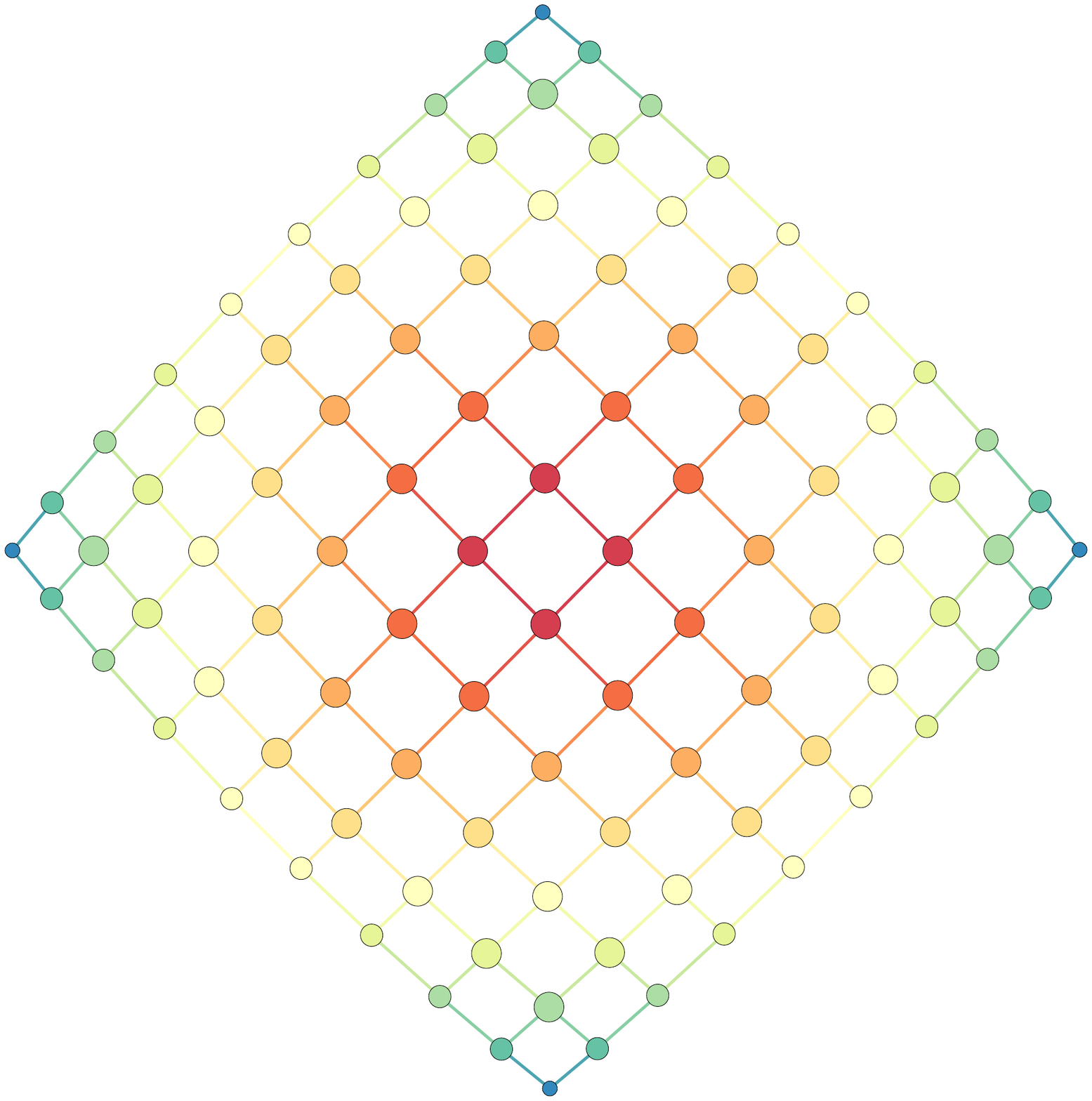}\\
    \includegraphics[trim=0.0cm 0.0cm 0.0cm 0.0cm, clip=true, width=0.32\linewidth]{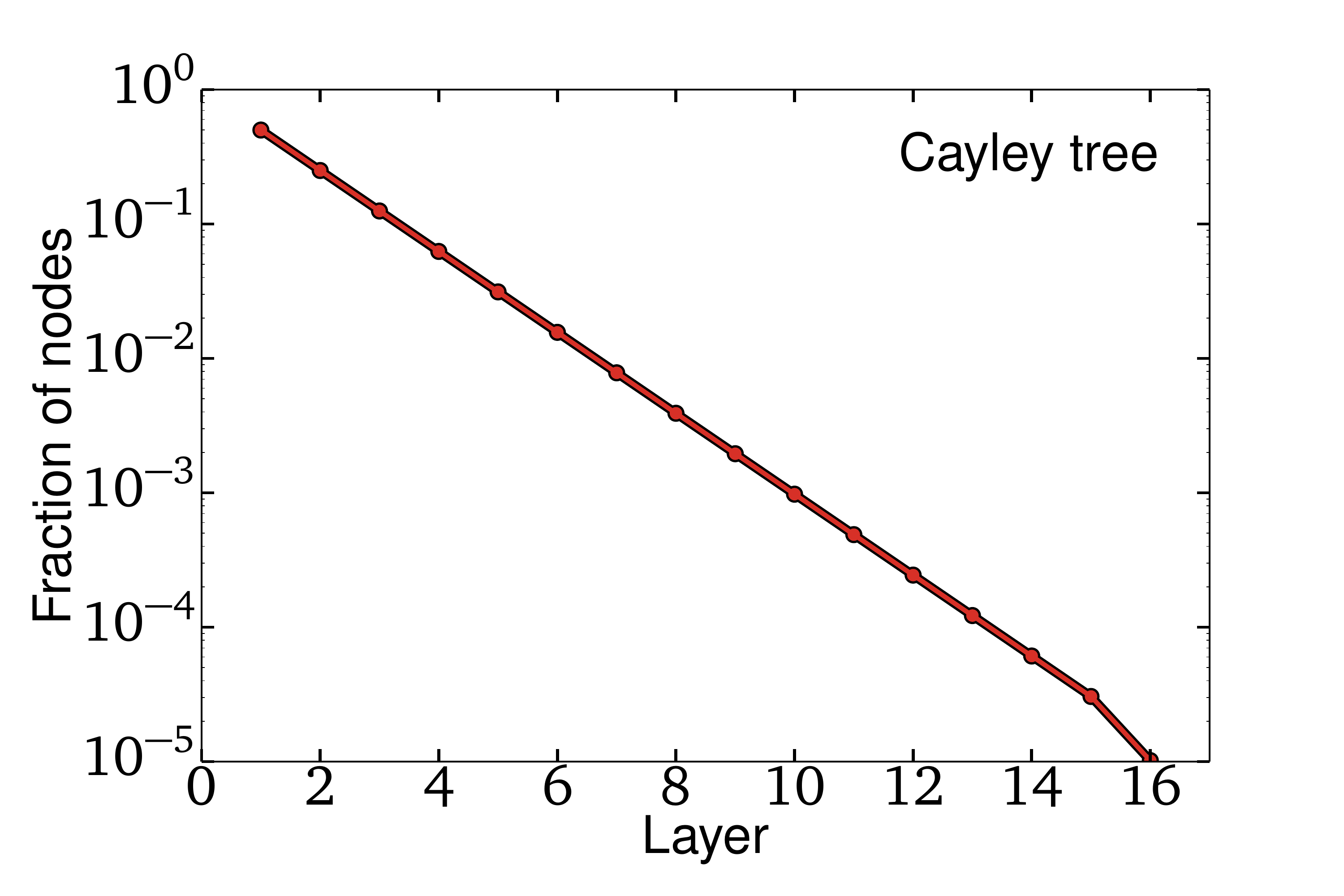} 
    \includegraphics[trim=0.0cm 0.0cm 0.0cm 0.0cm, clip=true, width=0.32\linewidth]{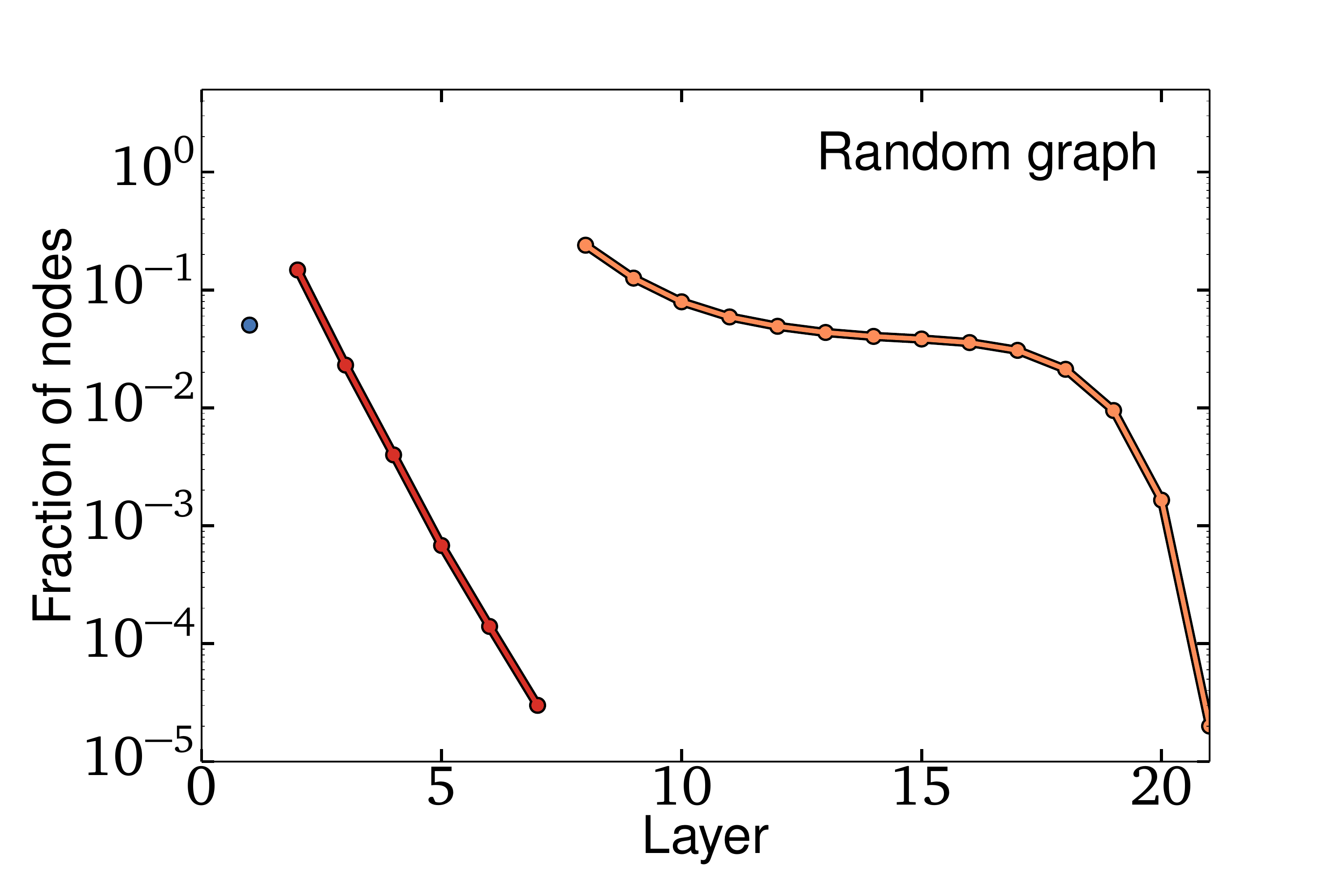} 
    \includegraphics[trim=0.0cm 0.0cm 0.0cm 0.0cm, clip=true, width=0.32\linewidth]{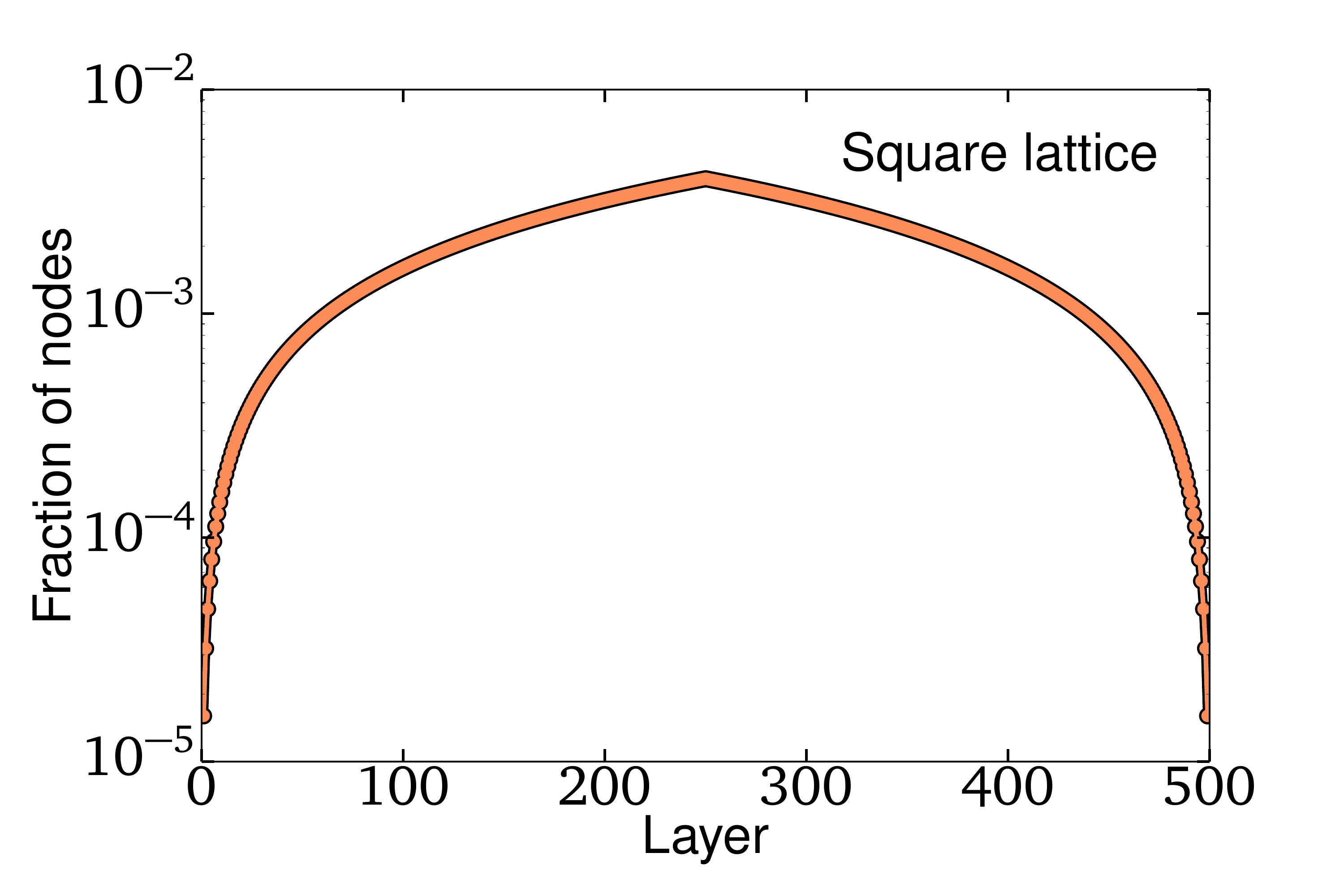}
    \caption{ \textbf{A few model networks and their onion spectra.} (left) The perfect tree with branching factor of two. (middle) A fully random Erd\H{os}-R\'{e}nyi graph. (right) A square lattice. In all cases, node size is proportional to its degree within the shown subset while node color indicates where the nodes are found in the onion decomposition, from blue for shallow nodes to red for deep nodes. In the onion spectra, dot colors and lines are used to indicate layers corresponding to the same $k$-shell. Note that colors are used for contrast and do not correspond to node color in the network cartoons. Both the Cayley tree and the square lattice only have one shell, corresponding to a $1$-core and a $2$-core, respectively, while the random graph features 3 cores, roughly corresponding to disconnected nodes, a tree-like periphery and a dense core. The distribution of nodes per layer in the Onion Decomposition tell us something about the structure as, for example, a tree is explored exponentially and a lattice linearly.
    }
    \label{fig:toynetworks}
\end{figure*}

The procedure to identify these layers, which we call the \textit{onion decomposition} (OD), is essentially the same as that for the $k$-core decomposition, but retains more information. We generalize the algorithm for $k$-cores \cite{Batagelj} with a very simple modification to produce the OD:

\algblock{If}{EndIf}
\algcblock[If]{If}{ElsIf}{EndIf}
\algcblock{If}{Else}{EndIf}
\algcblockdefx[Strange]{}{Eeee}{Oooo}
[1]{\textbf{Input} #1}
[1]{\textbf{Output} #1}
\begin{algorithmic}[1]
\Eeee{graph with vertex set $V$ and neighbors $N$}
\Oooo{Onion $Layer$ and $Coreness$ of each vertex}
\STATE{$D$ := list of degrees}
%\STATE{Sort $V$ and $D$ in ascending order by degree}
\STATE{$core$ := 1; $layer$ := 1}
\WHILE{$V$ is not empty}
\STATE{ThisLayer := $\{v \in V | D(v) \leq core\}$}
\FORALL{$v \in \text{ThisLayer}$}
  \STATE{$Coreness(v)$ := $core$; $Layer(v)$ := $layer$}
  \STATE{\textbf{for each} $w \in N(v)$, decrease $D(w)$ by 1 \textbf{end for}}
  \STATE{Remove $v$ from $V$ and $D$}
\ENDFOR
\STATE{$layer$ := $layer + 1$}
%\STATE{Sort $V$ and $D$ in ascending order by degree}
\IF{the minimum degree in $D$ is $\geq (core+1)$}
  \STATE{$core$ := $\min D$; $layer$ := 1}
\ENDIF
\ENDWHILE
\end{algorithmic}
The run-time of this algorithm scales as $O(|E| \log |V|)$. For the run-time analysis of the OD, as well as implementation notes that can affect the speed of the algorithm in practice, see Supplemental Material.

The OD thus provides a centrality measure more refined than coreness, and essentially at least as easy to compute as other centrality measures. For example, the best known algorithm for betweenness centrality takes time $O(|V||E| + |V|^2 \log |V|)$ \cite{Brandes2001}, and even computing eigenvector centrality takes time $O(|E|/\delta)$ (where $\delta$ is the gap between the first and second eigenvalues), which is comparable to the OD. In Section~\ref{sec:dis} we also show how the OD naturally corresponds to an ensemble of random networks, unlike other centrality measures. 

\section{Results}
With this new analysis method in hand, we define the \textit{onion spectrum} of a network as the fraction of all nodes which are found in a given layer of the OD. The onion spectrum can be thought of as a structural spectrum as it assigns every node to a given structural role through our new measure of node centrality. %(i.e., layer in the OD). 
As we will see, even a glance at the onion spectrum of different networks provides significant insights into their structure at multiple scales (see Table~\ref{table:corr}, whose entries are explained throughout the rest of the paper).

\begin{table}[h!]
\begin{center}
\caption{Summary of some properties found in the onion spectrum. ``Randomness'' here refers to rewiring preserving only degree distribution. All other properties can also be mapped to local connection rules using the random model presented in Sec.~\ref{one}.}
\begin{tabular}{ @{}c c c@{} } \hline
Scale\hspace{0.2cm} & Signature in the spectrum  & Property \\
\hline
\hline
\multirow{2}{*}{micro} & more cores than random & assortativity\\
& fewer cores than random & disassortativity\\
\hline
\multirow{3}{*}{meso} & exponential decay & tree-like structure\\
& sub-exponential decay & loopy structure\\
& change in decay & interesting subgraph\\
\hline
\multirow{2}{*}{macro} & cores & core-periphery structure\\
& layers & node centrality\\
\hline
\hline
\label{table:corr}
\end{tabular}
\end{center}
\end{table}

%%Power Grid and Penn roads
\begin{figure*}[ht!]
    \centering
    \includegraphics[trim=0.0cm 0.0cm -1.0cm 0.0cm, clip=true, width=0.4\linewidth]{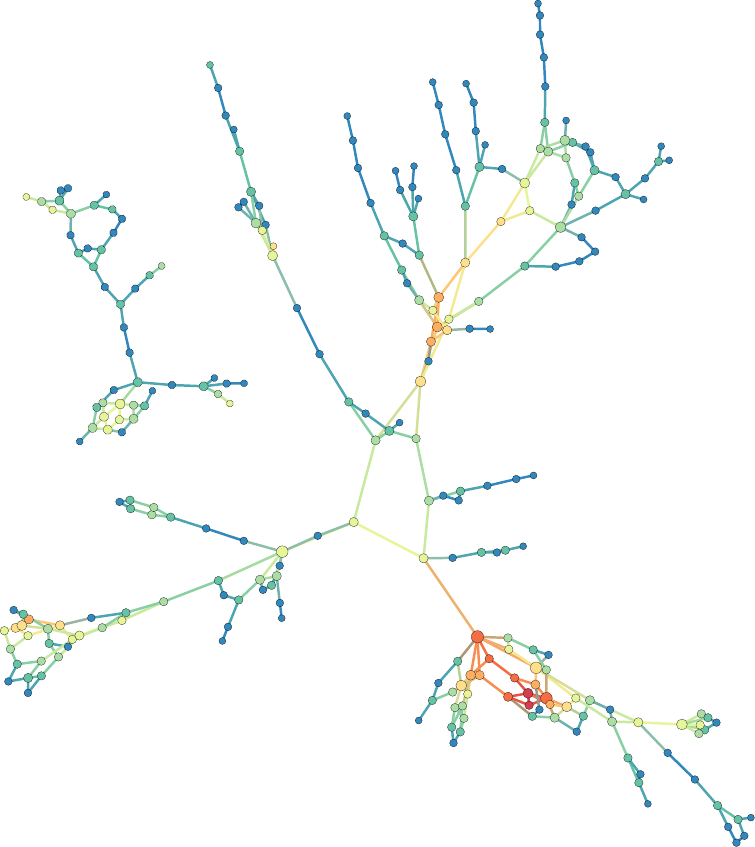}  
    \includegraphics[trim=0.0cm 0.0cm 0.0cm 0.0cm, clip=true, width=0.35\linewidth]{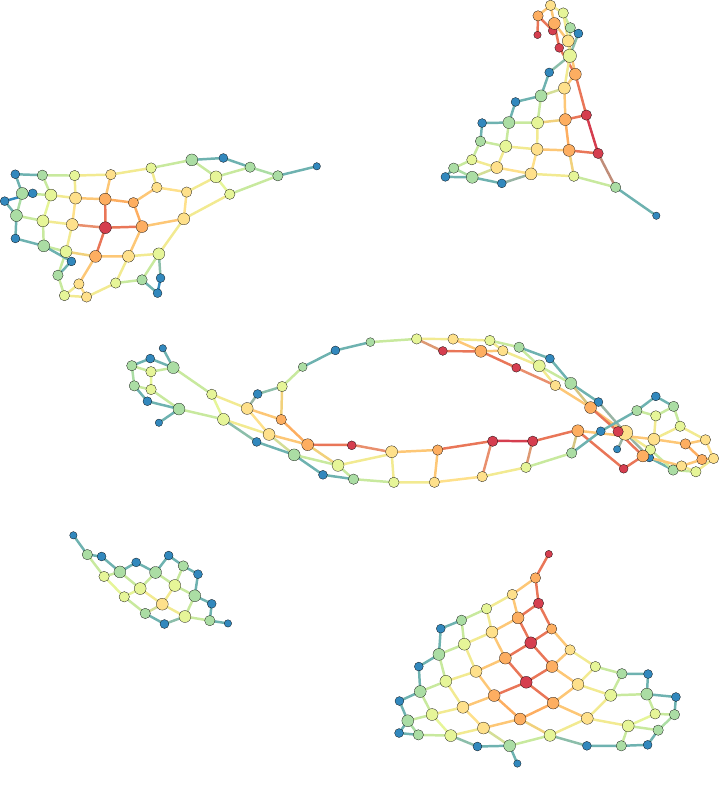}\\
    \includegraphics[trim=0.0cm 0.0cm 0.0cm 0.0cm, clip=true, width=0.45\linewidth]{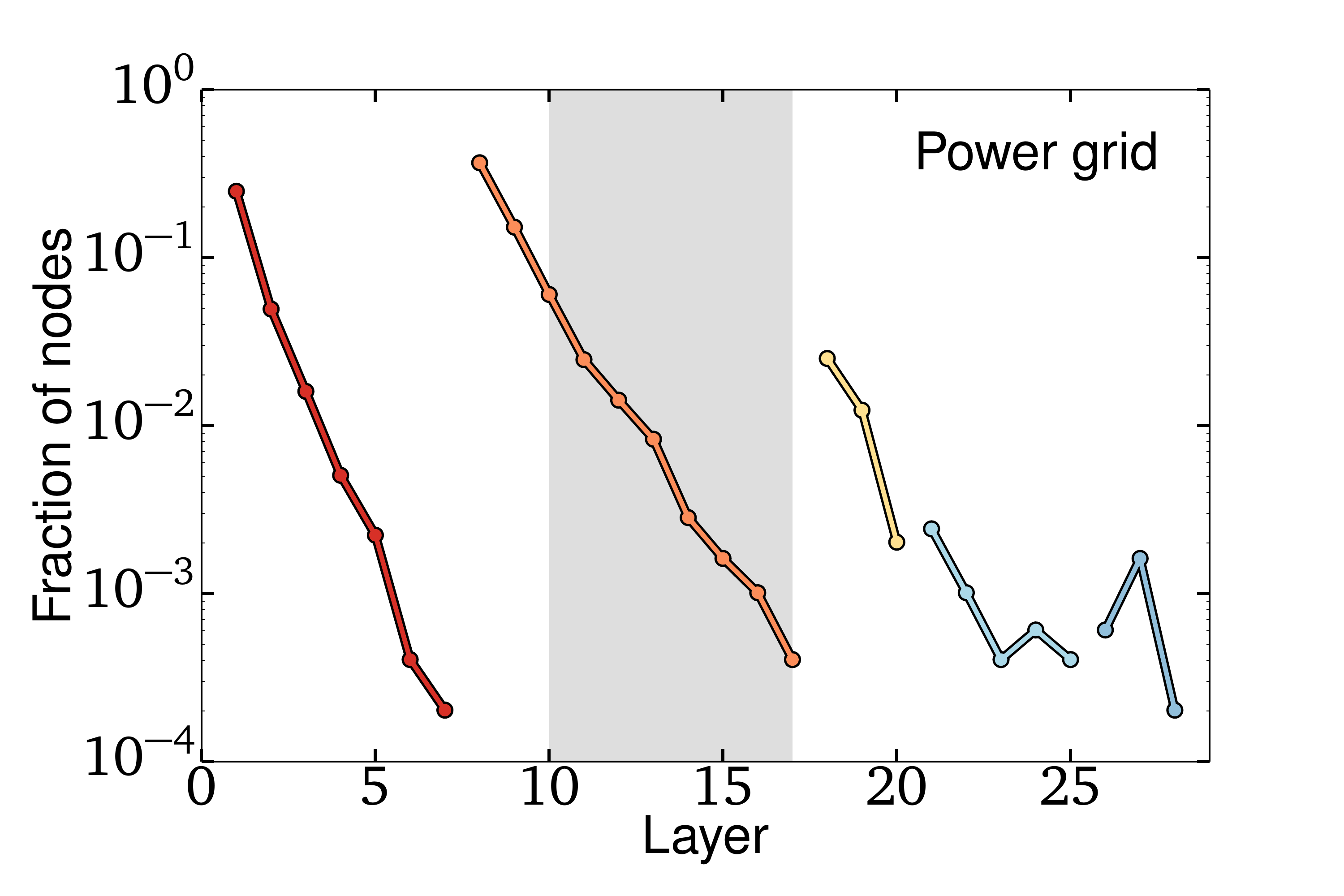}
    \includegraphics[trim=0.0cm 0.0cm 0.0cm 0.0cm, clip=true, width=0.45\linewidth]{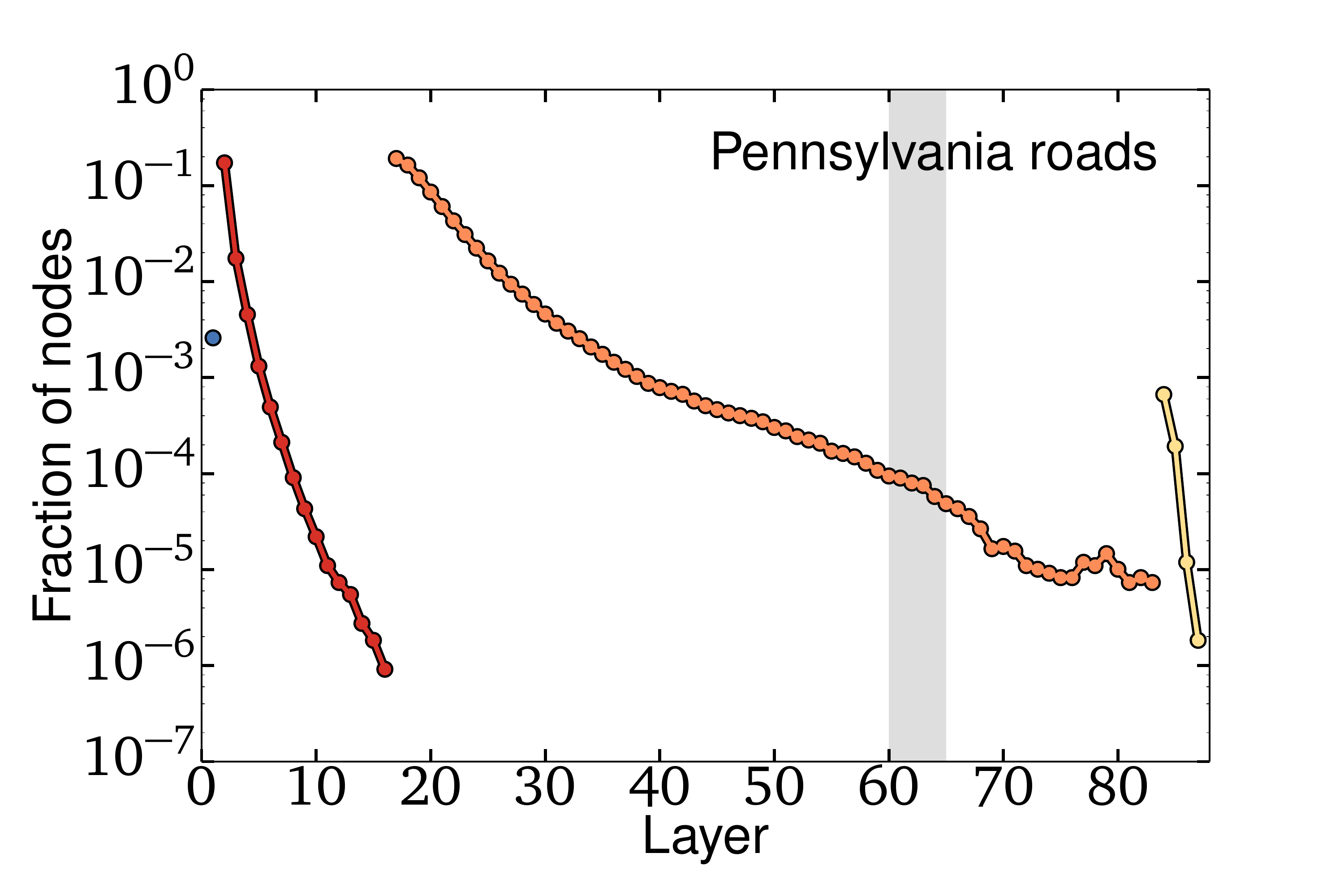}
    \caption{ \textbf{Onion spectra of the Northwestern American power grid and of the Pennsylvania road system and representative subgraphs.} As in the model networks of Fig. \ref{fig:toynetworks}, the tree-like structure of the power grid and the lattice structure of the road networks are reflected in the exponential and sub-exponential decay of layer density per shell in the onion spectra, respectively. Again, dot colors in the spectra reflect layers belonging to different shells whereas node color in the networks correspond to the shells in which they are found (shaded layers in the spectra, only links between nodes belonging to the subgraph are shown). 
    }
    \label{fig:intrastructure}
\end{figure*}

\subsection{Model networks}

We first test the OD on model networks designed to gather insights on the types of onion spectra produced by different network structures: A perfect Cayley tree with a branching factor of 2, an Erd\H{o}s--R\'{e}nyi graph with fixed density and a square lattice. Small versions of these networks are presented in Fig.~\ref{fig:toynetworks}. These networks are used to test the behavior of the OD on different structures: Tree-like branching, core-periphery structure and ``geographic-like'' embedding. We show the results of the OD on these model networks in Fig.~\ref{fig:toynetworks}. Certain features of their onion spectra can also be written analytically (see Supplemental Material).%, but it is also useful to compare these spectra to those of the rewired versions of the toy-networks, shown in Fig.~\ref{fig:rewiredtoynetworks}. We rewire the network using the Configuration Model which preserves the degree of each node and allow us to see how the actual networks differ from the most used random network model. 
\begin{figure*}[t!]
    \centering
    \includegraphics[trim=0.0cm 0.0cm 0.0cm 0.0cm, clip=true, width=0.45\linewidth]{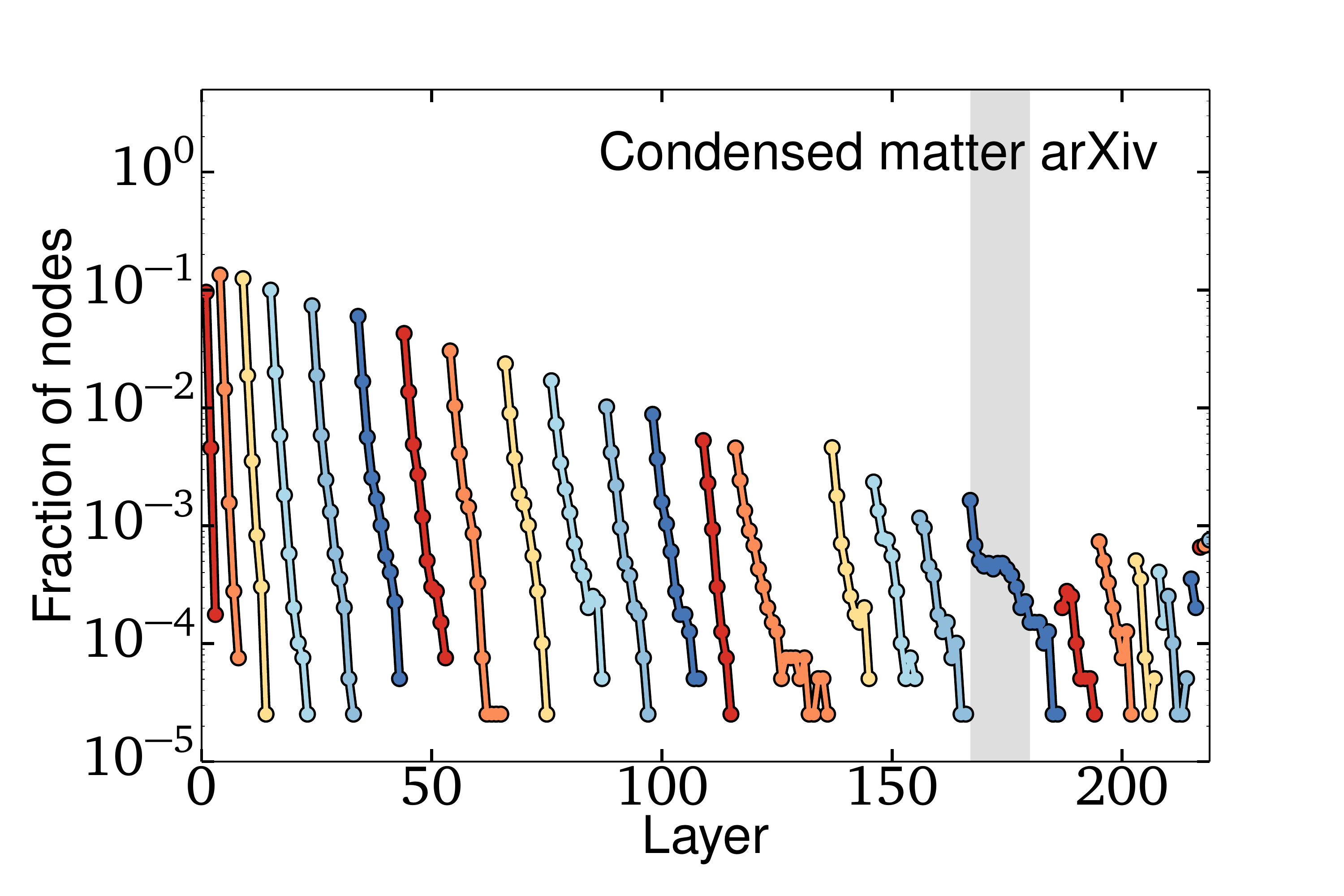}
    \includegraphics[trim=0.0cm 0.0cm 0.0cm 0.0cm, clip=true, width=0.45\linewidth]{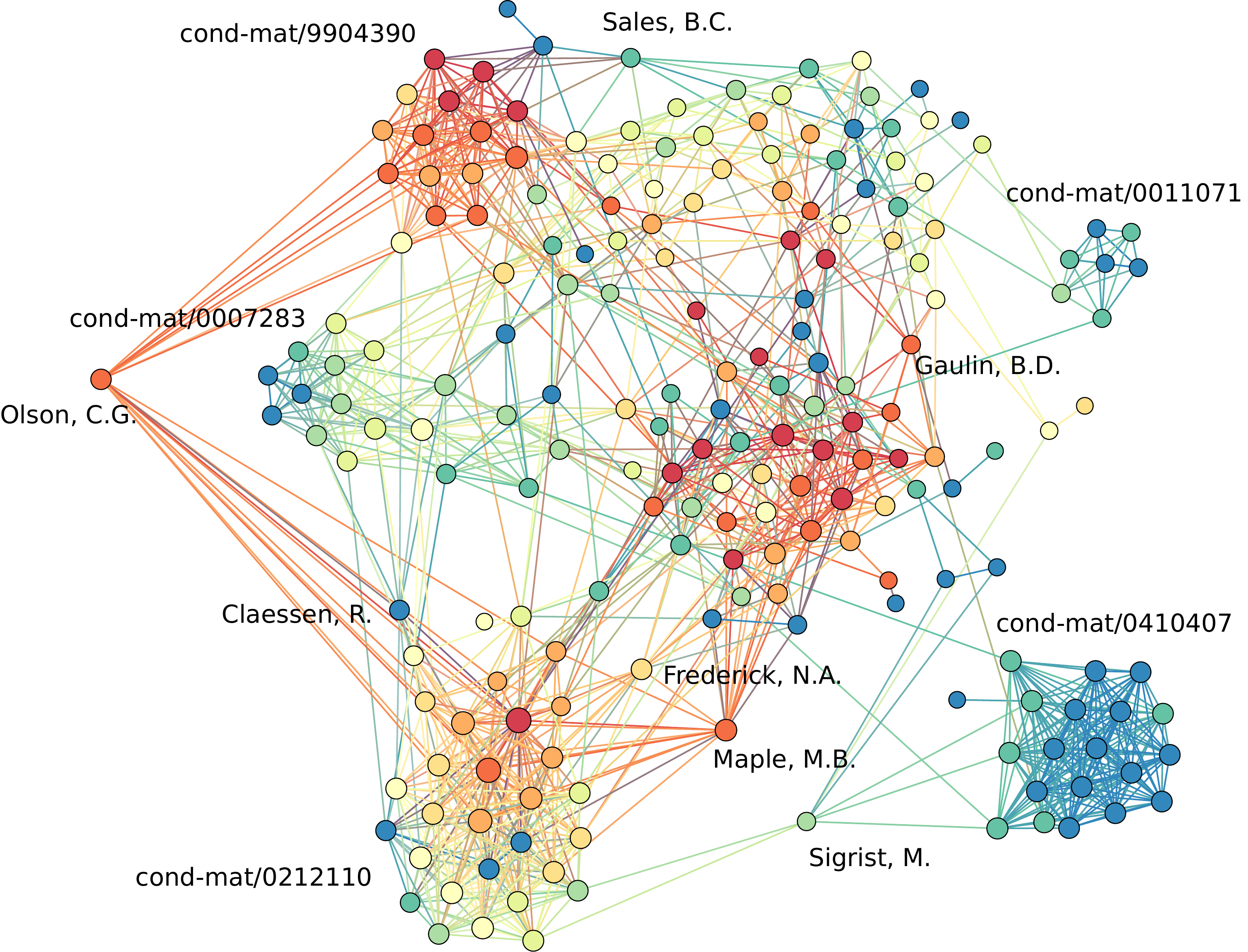}
    
    \caption{ \textbf{The condensed matter co-autorship onion spectrum and a selected subgraph.} While the overall structure of this sparse network is tree-like, some unique subsets can be identified in the spectrum, one of which is shown (shaded layers). Dense communities of collaborators are identified with a selected arXiv identifier and some of the authors bridging communities are identified by name.
    }
    \label{fig:condmat}
\end{figure*}

In the case of the Cayley tree, the nodes of the original network have a very well-defined position in the structure given by their distance to the leaf nodes. After removal of the first onion layer, the network is essentially the same (with one fewer layer) and the distances between all nodes and the leaves are reduced by one. The process then goes on, such that all nodes are within the $1$-core, and we can find an exponentially decreasing number of nodes in all layers (i.e., as an inverse branching process). In fact, all perfect trees feature an exponential onion spectrum where the ratio of two subsequent layer densities roughly corresponds to an average branching factor. Locally speaking, this tree-like property is observed in most sparse networks. % In the rewired network, all $N/2$ nodes of degree 1 are given the same role and all $N/2$ nodes of degree 3 are given the same role. By removing the strict constraints of the perfect tree, chances are that this rewiring process will now produce loops. Those loops force the appearance of $2$- or $3$-cores, thus removing nodes from the lower core and changing the purely exponential distribution of the perfect tree.

In the Erd\H{os}-R\'{e}nyi graph, the first few layers are peeled in a similar fashion to a perfect tree. This can be observed in Fig.~\ref{fig:toynetworks}(top center) where nodes of degree one (in blue) are connected to peripheral nodes of higher degree (in green). These various branches define the periphery of the network and comprise the first cores of the decomposition. The last core, much denser, features a significant number of loops on all scales. The density of onion layers within that core thus follow a very different decay, clearly sub-exponential until the finite size cut-off. To better understand this behavior, we turn to the square lattice, which possesses loops of all possible even lengths, by design.

%In the case of the Ravasz-Barab\'{a}si hierarchical network, nodes also have a well-defined structural role but their distance to the central node is no longer a good indicator of that role. In fact, this hierarchical network is self-similar in the sense that a base unit is repeated to obtain the full network. This implies that nodes of lower centrality are found everywhere in the network, including the neighborhood of very central nodes. Hence, the network breaks down very quickly under the OD as we change the degrees of many nodes at every pass. Moreover, the self-similarity of the network appears reflected in the similarity between the internal onion spectrum of the different cores.% In the rewired network, we lose both the fast break down of the network and the self-similarity between cores.

To analyze the square lattice, it is useful to go back to Fig.~\ref{fig:toynetworks}(top right) and notice how the network is explored by the OD. The first nodes to be removed are the corners, then their neighbours, and so on. The sites of nodes removed in each layer follow horizontal/vertical lines converging to the center with increasing layers. The unique shape of the onion spectrum is easy to calculate and is found to be linear, both in the increasing and decreasing regime (see Supplemental Material).% Once again, the rewiring process creates more core and changes the unique shape of the distribution; in this case, making the network much faster to decompose.

To broadly summarize the conclusions that can be drawn from the three model networks: The onion spectrum of tree-like and sparse random networks are expected to decay exponentially, whereas those of lattices and dense subgraphs fall sub-exponentially (see Table \ref{table:corr}).

\subsection{Real-world case studies}

We now apply the OD to several real-world networks. These networks are selected to cover a few structural features that are captured by the OD so as to highlight its use as a macroscopic, mesoscopic and microscopic tool (many more are explored in the Supplemental Material). Namely, we investigate how the onion spectrum allows us to distinguish tree-like and lattice-like networks, how it allows us to identify topologically anomalous subgraphs as features of interest, and how the degree distribution and degree-degree correlations are reflected in a network's onion spectrum (see Table~\ref{table:corr}). %Figure \ref{fig:allsprectra} presents the full spectrum of all considered networks. At a glance, a few interesting features can already be highlighted %difference between slopes across networks, but also across different regions of a given network.
\begin{figure*}[!ht]
    \centering
    \includegraphics[trim=0.0cm 1.5cm 0.0cm 0.0cm, clip=true, width=0.25\linewidth]{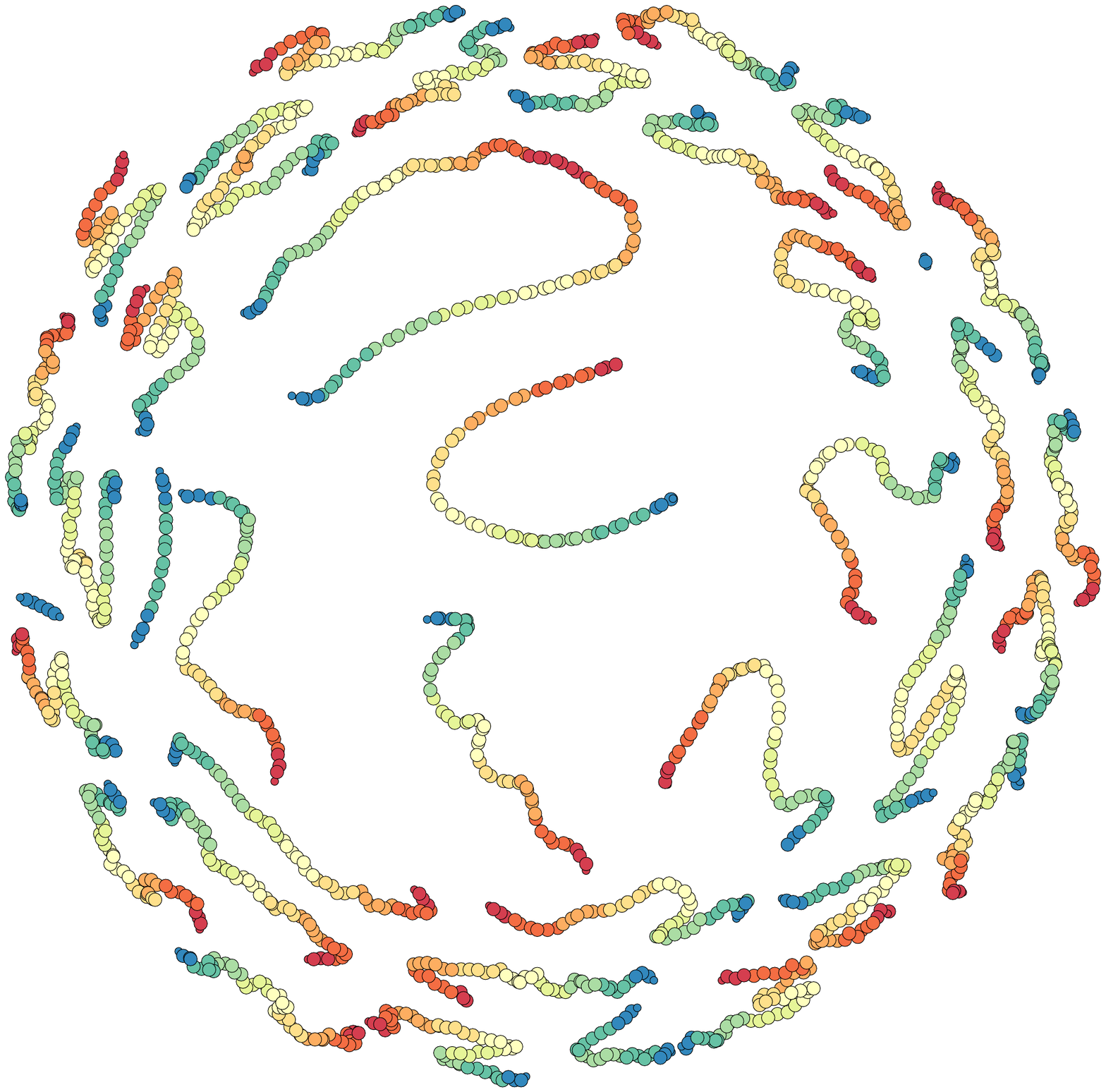}
    \includegraphics[trim=0.0cm 0.0cm 0.0cm 0.0cm, clip=true, width=0.45\linewidth]{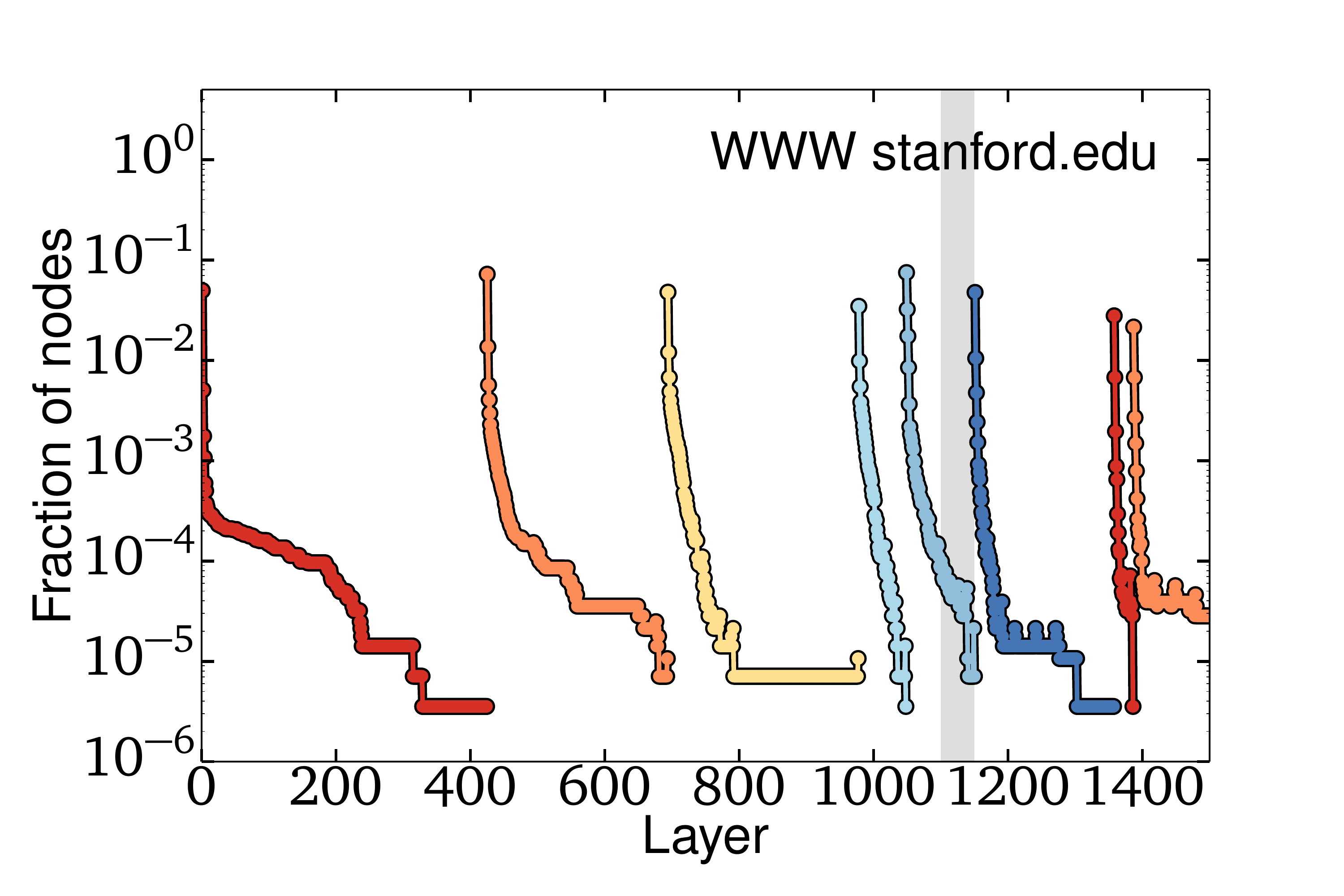}
    \includegraphics[trim=0.0cm -3.0cm 0.0cm 0.0cm, clip=true, width=0.25\linewidth]{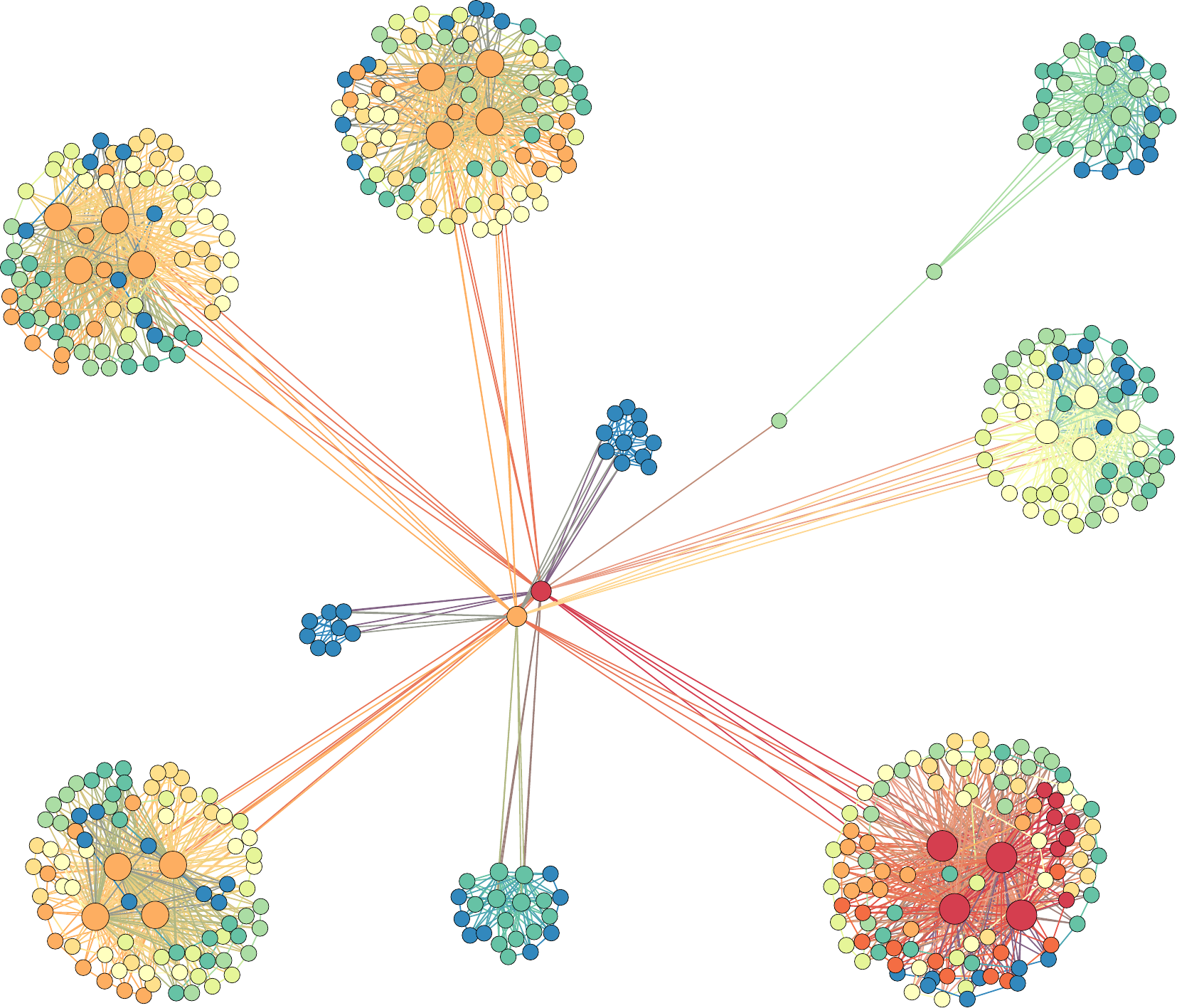}
    \caption{ \textbf{The WWW subset of \texttt{stanford.edu} and its onion spectrum.} This final case study highlights how the method can uncover backbones and governing nodes that can only be highlighted once peripheral nodes are removed. In particular, we illustrate this with extremely unlikely chains of nodes in the first core (left) and a centralized community structure in the sixth core (right).
    }
    \label{fig:wwwstanford}
\end{figure*}

%Networks with tree-like structure (power grids) or lattice-like (PennRoad)
Figure \ref{fig:intrastructure} presents the onion spectra of two infrastructure networks: The Northwestern American power grid \cite{Watts1998} and the Pennsylvania road system \cite{Leskovec}. In both cases, the figure presents a subset of the network that reflects its global structure. The selected layers are shaded in the spectrum, and parts of those subgraphs are presented in the top half of the figure. Once again, merely glancing at the decay of layer density informs us as to how these networks occupy space: The exponential decay of the power grid's onion spectrum is a signature of its (effectively) tree-like structure, where most loops are short and link nearby nodes. This is in stark contrast with the sub-exponential spectrum and structure of the road network, where despite featuring fewer triangles, there are loops occurring on all scales, as observed in Fig. \ref{fig:intrastructure}(top right). To visualize the difference, note that removing a link or two can disconnect many nodes from the power grid, but not on the road network.

%\begin{figure}[t!]
%    \centering
%    \includegraphics[trim=0.0cm 0.0cm 0.0cm 0.0cm, clip=true, width=0.69\linewidth]{Enron_pass_distribution} 
%    \includegraphics[trim=0cm -7.0cm 0.0cm 0.0cm, clip=true, width=0.3\linewidth]{Enron_core-crop}
%    \caption{ \textbf{The onion spectrum and dense core of the email network of Enron.} Bla bla ba
%    }
%    \label{fig:enron}
%\end{figure}

%Correlations in social networks
%Evaluating the rate of the exponential decay in the second core of the power grid predicts an average branching factor of 2.4. Indeed, the actual average excess degree of the power grid is 2.9 and the discrepancy corresponds to branching towards nodes of belonging to different cores. However, it should be noted that this evaluation works well because the decay rate remains relatively constant during the entire core. This is not always the case, and varying decay rates contain further information of the network structure.

While the major cores of both the power grid and the road network have a very steady behavior, one can look for significant deviation from general trends---i.e., topological anomalies---to identify interesting subgraphs that stand out from the overall network structure. To this end, we present the onion spectrum of a co-authorship network \cite{newman00_pnas} in Fig. \ref{fig:condmat}(left). Like most sparse social networks, the overall structure is roughly tree-like. However, the onion spectrum allows us to identify interesting subgraphs, leading us to focus on the nodes around layer 175 (highlighted), as we see a radically different decay there. This anomalous subgraph is composed of nodes with degree at least 18 that, once peripheral nodes are removed, are found to be organized in communities of long-time, prolific collaborators. % now much easier to identify. 
Similar topological anomalies, with similar explanations, are seen in other co-authorship networks (see Supplemental Material).
This locally high modularity contrasts with the global tree-like structure, and is easily picked up by the OD. A subset of this subgraph is presented on Fig. \ref{fig:condmat}(right). 

%Similarly, we can see the signal of a very dense subgraph around the 230th layer. The more loops a network contains, the more it is going to differ from a perfect tree and the more its spectrum will differ from an exponential decay. As we previously observed in the Erd\H{os}-R\'{e}nyi graph, very dense cores are an extreme case of this behavior, as seen in this network in Fig. \ref{fig:mathsci}(bottom).

The condensed matter co-authorship network also illustrates how the onion spectrum can detect degree correlations. %the decay rate can differ from layer to layer. 
Not only does this network have more cores than the previous networks because of its fat-tailed degree distribution, it even has significantly more cores than would be expected from a network with the same degree distribution but otherwise random connections: 29 versus 9, respectively (see Supplemental Material).
%This network has more cores than the previous networks, simply because of its fat-tailed degree distribution; however,  Yet, we also see significantly more cores than we would have expected from a network with the same degree distribution but otherwise random connections: 29 versus 9, respectively.  
This is a signature of positive degree correlations, often called \textit{homophily} \cite{mcpherson} or assortativity, meaning neighboring nodes tend to have similar degrees \cite{newman2003}. It is intuitively clear that having high-degree nodes share connections with other high-degree nodes favors the emergence of higher coreness. By comparison to a randomized network, one can thus look for degree correlations by comparing how many cores are found within the network (see Supplemental Material for more examples). That being said, the decay rates within cores can also contain further information about the network structure and this is especially true when faced with negative degree correlations.

%Now that we have established the signature of homophily, we can look at a classic case of negative degree correlations. 
The onion spectrum of a subset of the World Wide Web, the \texttt{stanford.edu} domain \cite{Leskovec}, is presented in Fig. \ref{fig:wwwstanford}(middle). In this network, even within a single layer, the decay rate varies significantly. In the Supplemental Material, we show that this variation is a consequence of negative degree-degree correlations, by comparing the network with two ensembles: One in which only the degrees are preserved, and one in which furthermore the degree-degree correlations are preserved. In the former, the decay rates within each layer are roughly constant, whereas in the latter the decay rates closely match the true, observed rates. 
%Fig. \ref{fig:myspace} \cite{Ahn2007}. Even within a single layer, the decay rate varies significantly. To illustrate how this is an effect of network structure and not simply an artefact of the degree distribution, we also present the onion spectrum of a rewired Myspace network while preserving every node's degree. Yet, if we simply also preserve degree-degree correlations between pairs of neighboring nodes \cite{Newman2002}, we recover the initial large gap in decay rate. 
Essentially, this variation is caused by the fact that most links stemming from low degree nodes are connected to nodes of higher degree than would be expected by chance. This negative correlation, or disassortativity, implies that fewer nodes are removed within the second and subsequent layer of a core than expected. This also implies that the network contains fewer cores than a randomized version. We propose a better rewiring scheme in Sec. \ref{sec:dis}.

%\begin{figure*}
%    \centering
%    \includegraphics[trim=0.0cm 0.0cm 0.0cm 0.0cm, clip=true, width=0.32\linewidth]{Myspace_pass_distribution} 
%    \includegraphics[trim=0.0cm 0.0cm 0.0cm 0.0cm, clip=true, width=0.32\linewidth]{CMofMyspace_pass_distribution} 
%    \includegraphics[trim=0.0cm 0.0cm 0.0cm 0.0cm, clip=true, width=0.32\linewidth]{CCMofMyspace_pass_distribution} 
%    \caption{ \textbf{The onion spectra of the Myspace online network and of two randomized versions.} Randomizing a network for comparison can highlight some of its properties. In this case, the total number of cores and the density of their secondary layers in the Myspace network (left) are a sign of its negative degree correlations. This is confirmed by comparing the spectrum of Myspace to that of randomized version which remove (middle) or preserve (right) the degree correlations. These three spectra were stopped at the 150th layer to highlight the initial behaviour.
%    }
%    \label{fig:myspace}
%\end{figure*}
%%WWWStanford

Finally, we use the \texttt{stanford.edu} domain to illustrate how other types of interesting subgraphs can be identified. %(Many more real-world networks are examined in the Supplemental Information.) 
In the case of most sparse networks, being ``interesting'' means at least deviating from the expected tree-like structure. In Fig. \ref{fig:wwwstanford}, we identify two (out of many) interesting subgraphs that present a significantly sub-exponential decay of density per layer. The first is interesting as it appears in the first core, where nodes are expected to be of low degree such that large non-tree-like subgraphs are unlikely. Yet, through the OD, we have identified a set of around 8500 nodes of degree two joined in chain-like fashioned (approximatively 3\% of the entire network). Based on the degree distribution alone, these subgraphs had a probability less than $10^{-15000}$ of occurring (i.e., it should not occur). But somehow, such surprising structures seem to appear on all scales within the domain, as similar slow decay can be observed in all cores. For instance, the subgraph shown in Fig. \ref{fig:wwwstanford}(right) occurs within the 5-core and appears to be a centralized community structure where many different groups are governed by only a few nodes. The entire subgraph collapses under the OD once the smallest groups are removed. In the context of World Wide Web networks, we thus believe that this procedure can help uncover the backbone and governing nodes of different domains.

\begin{figure*}[ht!]
    \centering
    \includegraphics[trim=0.0cm 0.0cm 0.0cm 0.0cm, clip=true, width=0.45\linewidth]{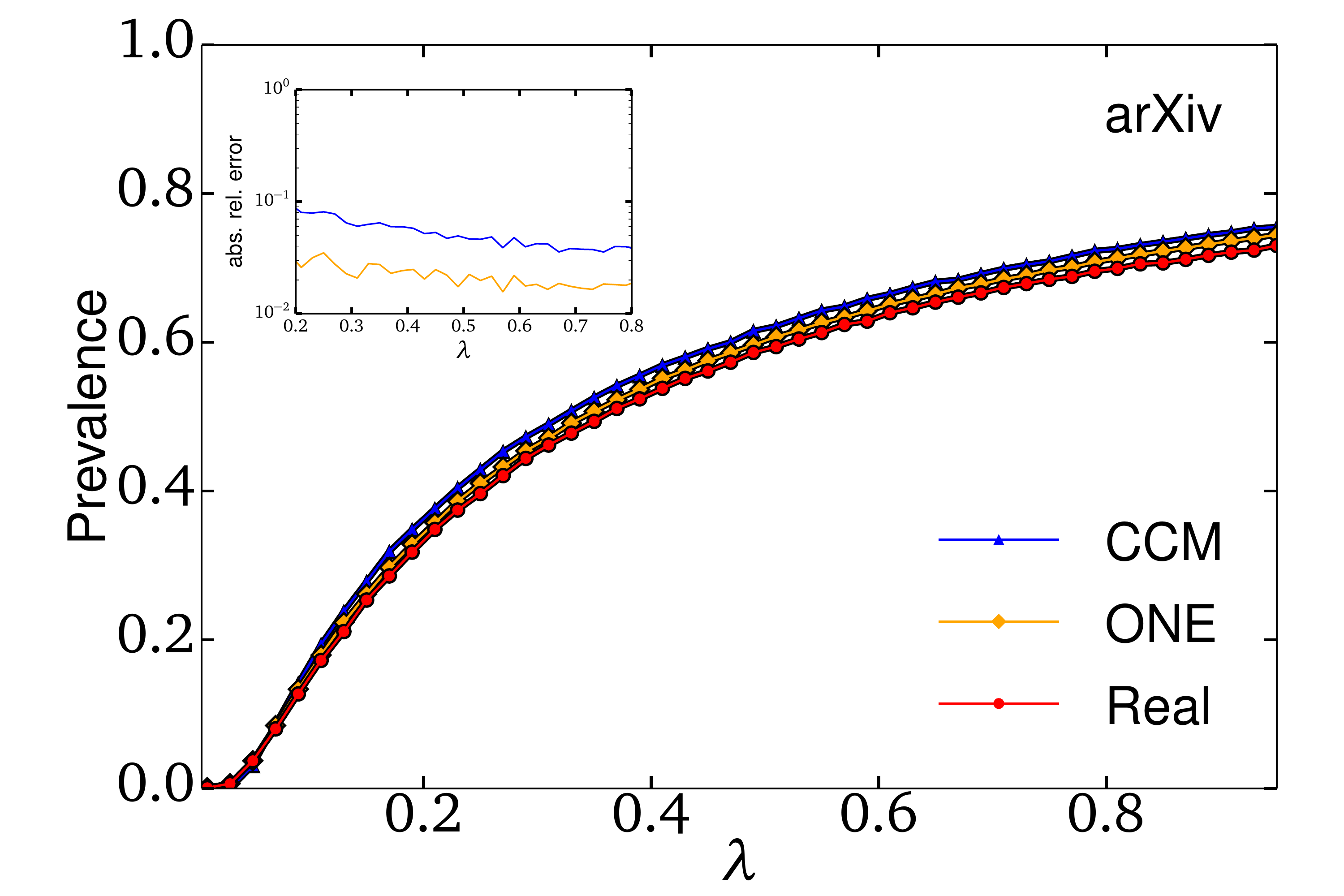}
    \includegraphics[trim=0.0cm 0.0cm 0.0cm 0.0cm, clip=true, width=0.45\linewidth]{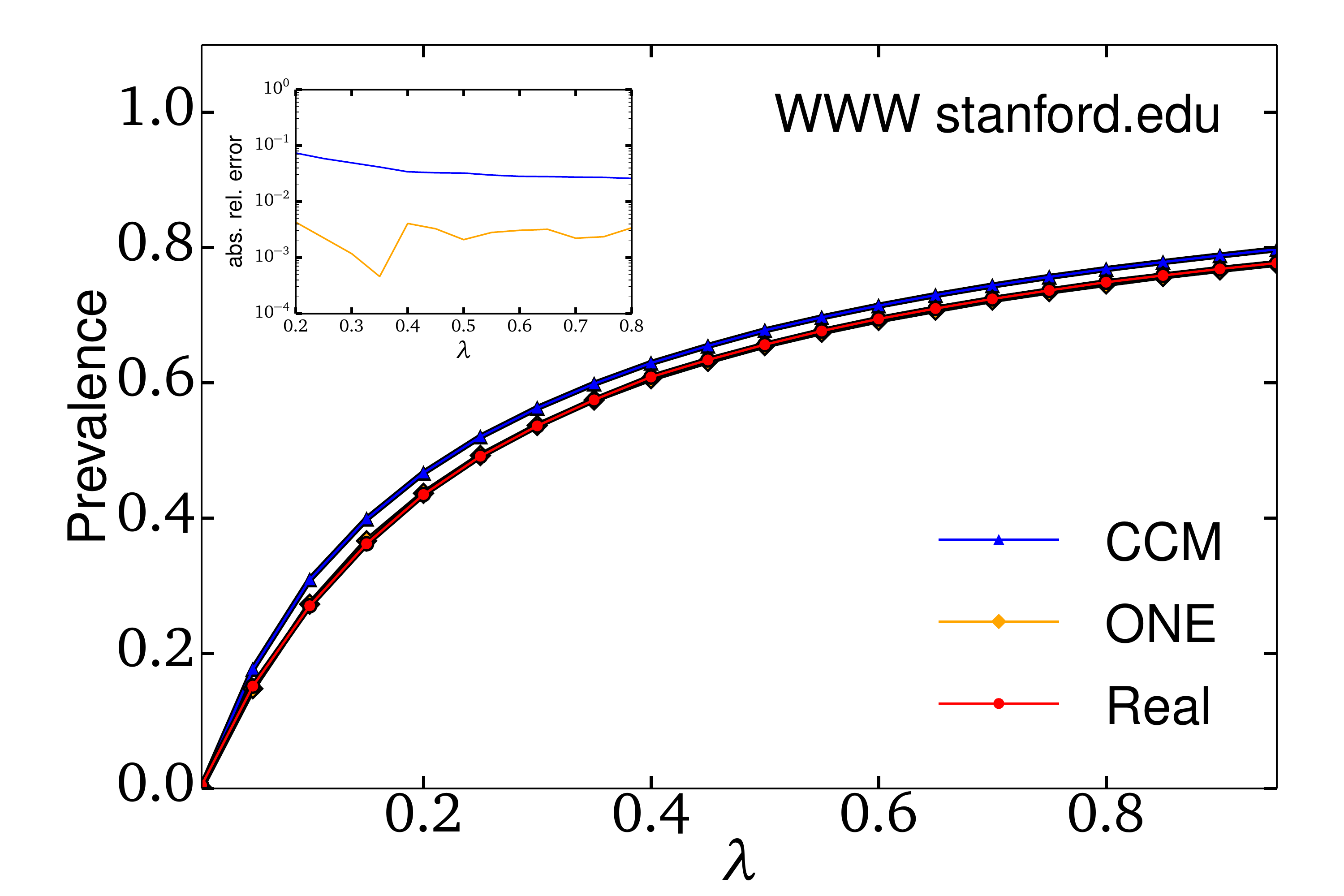}
    \caption{ \textbf{Simulations of the SIS epidemic process.} The prevalence (number of infectious nodes) associated with the steady state of the SIS dynamics as a function of the dimensionless parameter $\lambda$ on (left) the co-authorship network from the arXiv and (right) links between \texttt{stanford.edu} pages. Simulations on the real original networks are shown in red, and those on networks rewired according to the Correlated Configuration Model (CCM) and the ONE are shown in blue and orange, respectively. The insets show the absolute value of the relative errors on predictions from the rewired networks. Note that prevalence observed on the \texttt{stanford.edu} network and on its ONE are almost indistinguishable. In addition to its more accurate results, the ONE requires significantly less information than the CCM.}
    \label{fig:sis}
\end{figure*}

Although other methods might be designed to detect these particular structures---backbones, governing nodes, and the structure in Fig.~\ref{fig:condmat}---we point out that the OD allowed us to detect these structures in these networks \emph{de novo}, \emph{without} knowing what kind of structure we were looking for ahead of time.

% =================================================================================================
\section{Discussions and applications\label{sec:dis}}
% =================================================================================================
%
We have shown how the onion decomposition allows to characterize a complex network, at a glance, through the general trends of its onion spectrum. The OD achieves this in part because it aggregates multiple properties of complex networks, at different scales, into a single spectrum: Degree heterogeneity and correlations, organization in terms of centrality, and tree-likeness or loop prevalence.
%
%We have thus far demonstrated how the OD aggregates multiple properties of complex networks into a single spectrum: node heterogeneity, centrality structure, degree correlations, and tree-likeness or loop prevalence. The OD thus allows us to characterize a complex network, at a glance, through the general trends of its onion spectrum. Moreover, focusing on layers that go against those general trends allows us to identify uniquely interesting subgraphs within the network. For example, we can find large communities within an otherwise random network or a sparse periphery to an otherwise dense network.

To the best of our knowledge, the onion decomposition is the first method to combine so many potential tools into one algorithm that is simultaneously fast (it scales as $O(|E| \log |V|)$), simple, well-defined, mathematically principled, and allows \emph{de novo} detection of interesting subgraphs. Indeed, and contrary to other methods of network analysis (e.g., community detection algorithms), the positions of the nodes in the OD are elegant and straightforward: (i) a node is in the $k$-core if it is in the maximal subset of the network in which every node has degree at least $k$ within the subset, and (ii) within the $k$-core, a node is in the $l$-th local layer if one must remove all nodes of degree at most $k$ a total of $l-1$ times before reaching the node.
%Indeed, the proposed algorithm is very simple and scales as $O(|E| \log |V|)$. 
%
% \begin{enumerate}
% \item A node is in the $k$-core if it is in the maximal subset of the network in which every node has degree at least $k$ within the subset;
% \item Within the $k$-core, a node is in the $\ell$-th local layer if one must remove all nodes of degree at most $k$ a total of $\ell-1$ times before reaching the node.
% \end{enumerate}

%Finally, we show how the OD naturally leads to the most structurally constraining, yet still easily and exactly sampleable, network randomization procedure to date.
As a final point, we show how the OD naturally allows to define the most structurally constraining, yet still easily and exactly sampleable, network randomization procedure to date. We then show how these constrained networks outperform existing random network models at reproducing the outcome of the Susceptible-Infectious-Susceptible dynamical process, as well as at reproducing the full distribution of all-pairs shortest path lengths.
%
%We then show how these constrained networks outperform existing random network models at reproducing the outcome of the Susceptible-Infectious-Susceptible dynamical process.
%
%
%
%
% =================================================================================================
\subsection{The onion network ensemble (ONE)\label{one}}
% =================================================================================================
%
% Many properties of networks can be efficiently summarized by the joint degree-onion spectrum, i.e., each node gets assigned its degree and its location within the onion spectrum. This naturally leads to the definition of an ensemble of networks that preserve the joint degree-onion spectrum, which we refer to as the onion network ensemble (ONE). 
%
The specifications of the OD can be translated into a simple set of connection rules defining an ensemble of networks that preserve the joint degree-onion spectrum (i.e., the degree and the position in the onion spectrum of nodes). We refer to this ensemble as the onion network ensemble (ONE). Given a joint degree-onion spectrum, the ONE can be sampled quasi-uniformly, by a model only marginally more complicated than the configuration model. This is in strong constrast with other centrality measures such as betweenness or eigenvector centrality, whose translations into local connection rules, if any, remain open problems.
%To sample from the ONE, we show how to map the onion structure onto local pairwise rules of connection. 

In the ONE, a node of degree $d$ and coreness $k$ belonging to layer $\ell$ must be connected to other nodes in the following fashion. If the node is located in the first layer of its core (i.e., nodes in layer $\ell-1$ have coreness $k-1$), it must have exactly $k$ links to nodes of layers $\ell' \geq \ell$. % and $d-k$ links to nodes of layers $\ell' < \ell$. 
Otherwise, the node must have at least $k+1$ links to nodes of layers $\ell' \geq \ell-1$ and at most $k$ links to nodes of layers $\ell' \geq \ell$. %; the remaining of the links are to nodes of layer $\ell'<\ell$.
Rewiring links according to these rules while enforcing the correlations between layers \footnote{The fraction of links between any given two layers of a given original network is preserved on average. See \cite{lhd_2013_pre} for an example of the procedure.} allows to explore the ONE associated to a given original network, while simultaneously preserving its degree distribution and its onion spectrum.
%
% A node of degree $k$ and coreness $c$ belonging to layer $\ell$ must be connected to other nodes in the following fashion: $c$ of its $k$ links must go to nodes of layers $\ell' \geq \ell-1$ while its remaining $k-c$ links must go to nodes of layers $\ell' < \ell$. Additionally, if the node is located in the first layer of its core, all $c$ of its contributing links must be to layers $\ell' \geq \ell$. Otherwise it must have at least one link (contributing or not) with nodes in layer $\ell' = \ell-1$ to assure its position in the OD. Rewiring links according to these rules while enforcing the correlations between layers \cite{lhd_2013_pre} allows exploring the ONE associated to a given network, which simultaneously respects its degree distribution and its onion spectrum.
% 
% This rewiring process can also be used to build networks \emph{of different sizes} with the same joint degree-onion spectrum. This allows comparison of graphs across sizes, as well as analysis in the limit of infinite size.
This rewiring process provides a structurally constrained null model against which real networks can be compared, and that can also be used to build networks \emph{of different sizes} with the same joint degree-onion spectrum. This allows comparison of graphs across sizes, as well as analysis in the limit of very large networks.
%
% Given a joint degree-onion spectrum, the ONE can be sampled perfectly uniformly, by a model barely more complicated than the configuration model, unlike other centrality measures such as betweenness or eigenvector centrality. To sample from the ONE, we show how to map the onion structure onto local pairwise rules of connection. A node of degree $k$ and coreness $c$ belonging to layer $\ell$ must be connected to other nodes in the following fashion: $c$ of its $k$ links must go to nodes of layers $\ell' \geq \ell$ (links contributing to its coreness) while its remaining $k-c$ links must go to nodes of layers $\ell' < \ell$ (links not contributing to its coreness) with at least one in layer $\ell' = \ell-1$ (anchor link to assure its position in the OD). The first two types of links can then be chosen to respect the correlations between layers \cite{lhd_2013_pre}. Any network can thus be mapped to (and approximated by) an infinite network that simultaneously respects its degree distribution and its onion spectrum.
%
%We thus propose that networks can be efficiently summarized by the joint distribution, for each node, of its degree and location within the onion spectrum. We refer to the network model specified by this joint distribution as the onion network ensemble (ONE). This model is motivated by our results above and by the two following consequences. 
%
Note that in this limit, the ONE exactly defines a network without loops or correlations (other than the ones between layers that are explicitly enforced). %The second statement then follows from the first.%, if we assume that the network has no loop, we can construct an infinite tree that follows any set of degree distribution and onion spectrum.
\begin{figure*}[ht!]
    \centering
    \includegraphics[trim=0.0cm 0.0cm 0.0cm 0.0cm, clip=true, width=0.45\linewidth]{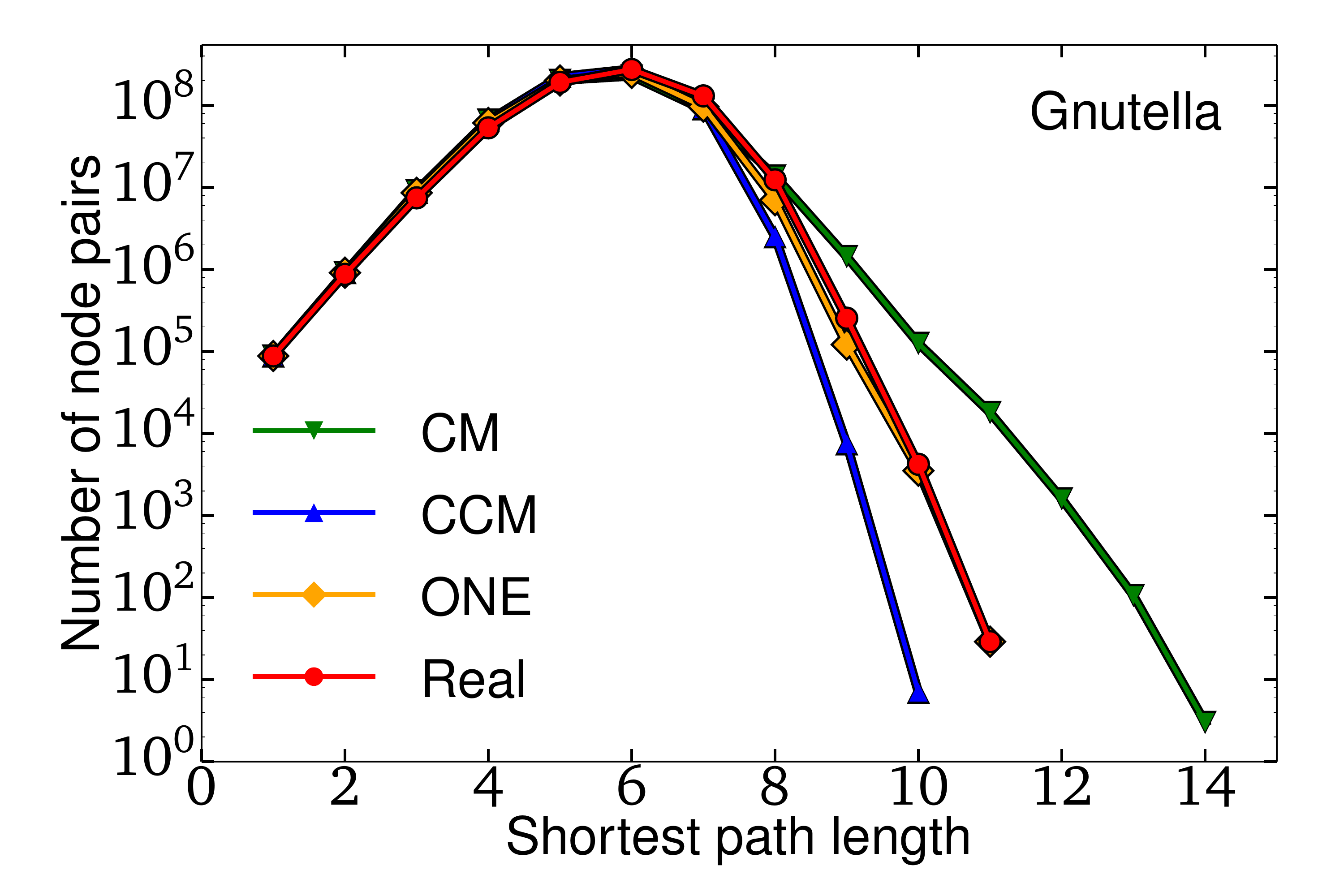}
    \includegraphics[trim=0.0cm 0.0cm 0.0cm 0.0cm, clip=true, width=0.45\linewidth]{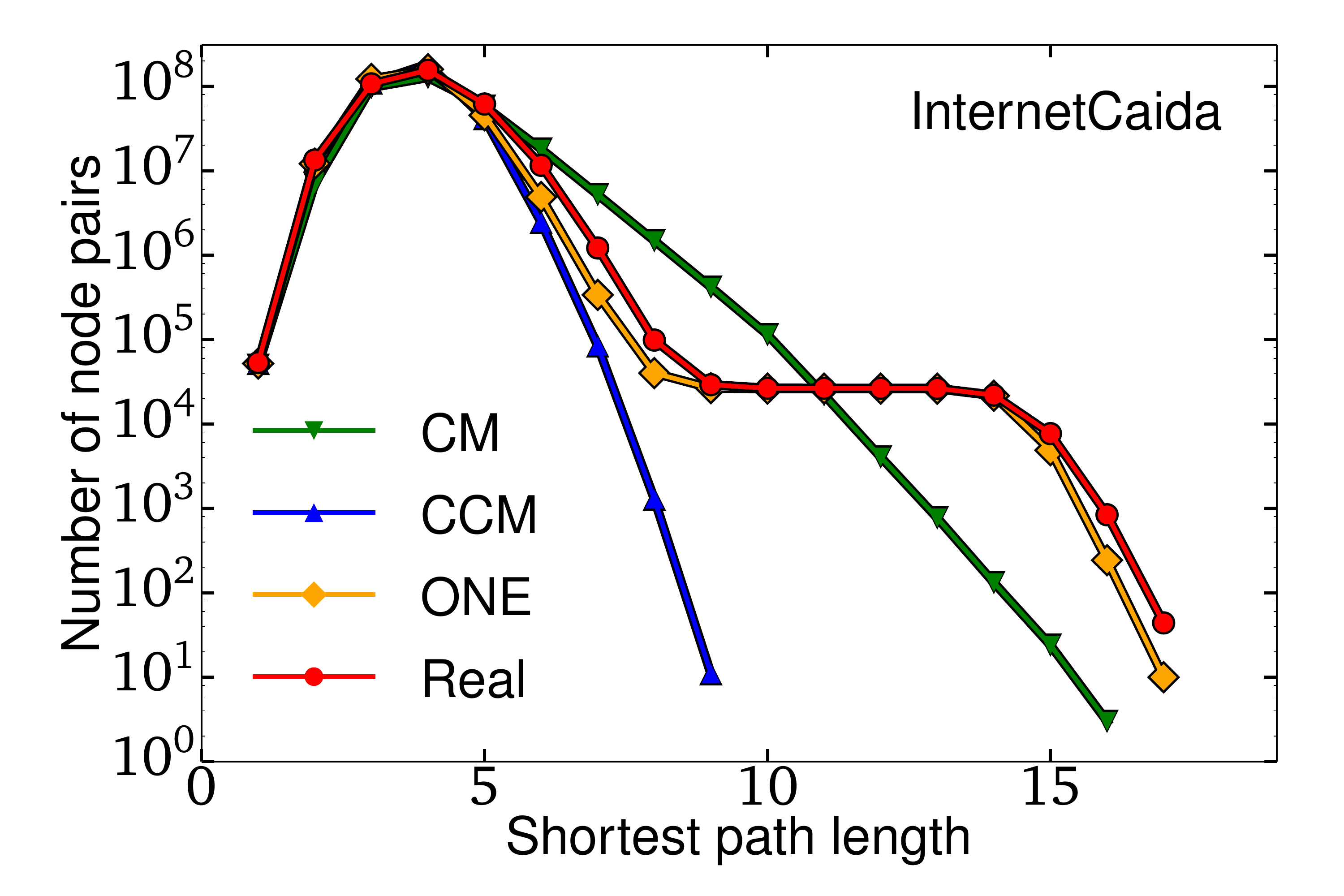}
    \caption{ \textbf{Distribution of the length of the shortest paths.} The distribution of the length of the shortest path between all pairs of node in (left) the Gnutella peer-to-peer sharing network \cite{lhd_2013_pre} and (right) the Internet (circa 2004-2007) \cite{leskovec05}. The distributions of the length of shortest paths in the original networks are shown in red, and the distributions for the rewired networks according to the Configuration Model (CM), the Correlated Configuration Model (CCM) and the ONE are shown in green, blue and orange, respectively. Note that for the Internet, the CM and CCM don't even get the \emph{shape} of the distribution correct, whereas the ONE gets not only the shape correct, but very nearly matches the exact values.}
    \label{fig:paths}
\end{figure*}
%
% =================================================================================================
\subsection{Dynamical process on the ONE\label{sis}}
% =================================================================================================
%
Let us consider the Susceptible-Infectious-Susceptible model of disease spread as an illustration of the general utility of the OD---and more specifically of the ONE---as an effective network structure. In this dynamics, nodes are either susceptible or infectious, infectious nodes infect their susceptible neighbors at a rate $\beta$ and they recover at a rate $\alpha$. The SIS dynamics is governed by the dimensionless parameter $\lambda = \beta / \alpha$ and, for $\lambda$ above a critical value, possesses a steady state in which a non-zero fraction of the nodes are infectious \cite{anderson91}. %infectious nodes which depends solely on the dimensionless parameter $\lambda = \beta / \alpha$.
% We now use a simple dynamical process to demonstrate that, despite being a lighter model, the ONE does a better job at capturing network structure. The process is question is the Susceptible-Infectious-Susceptible model of disease spread \cite{anderson91}. Infectious nodes recover (and become susceptible) at a rate $\beta$, but also infect each of their susceptible neighbours (which then become infectious) at a rate $\alpha$. The process leads to a steady-state fraction of infectious nodes which depends solely on the dimensionless parameter $\lambda = \beta / \alpha$.
%

We simulate this process on two real networks on which diseases, ideas or computer viruses could be expected to spread. Figure~\ref{fig:sis} gives the prevalence of the SIS dynamics as a function of $\lambda$ and compares it with the one observed on networks rewired according to the correlated configuration model (CCM) and the ONE. The CCM produces networks that respect a given degree distribution as well as the entire degree-degree correlation matrix (i.e., probability that a randomly selected link connects nodes of degree $d_i$ and $d_j$). The most striking result of Fig.~\ref{fig:sis} is that the ONE outperforms the CCM even though it requires \textit{less} information. Indeed, the CCM requires a joint degree-degree correlation matrix which scales as $k_{\textrm{max}}^2$, whereas the ONE requires information that scales as $k_{\textrm{max}}\ell_{\textrm{max}}\sim k_{\textrm{max}}^{3/2}$ (see Supplemental Material). This comparison with the CCM reveals that the performance of the ONE is not merely a consequence of the degree-degree correlations encoded in the layer-layer correlations, but stems from the global features encoded in the OD and in the ONE's local connection rules.
%a consequence that the local rules of connection of the ONE encode information about the global organization of networks.

It is worth noting that the SIS dynamics is highly self-averaging, as the infection constantly moves and revisits every node. Consequently, the SIS model does not depend much on the high-level structure of a network, but more on local features. Considering how much \textit{local} information is needed by the CCM, it is somewhat surprising that the ONE also performs better than the CCM in an application that depends mostly on local information.

%
% =================================================================================================
\subsection{Structure of the ONE\label{shortest_paths}}
% =================================================================================================
%
An interesting feature of our SIS results presented in Fig.~\ref{fig:sis} is that the ONE manages to provide precise approximations despite being a tree-like random network model. Indeed, it is often thought that models like the CCM overestimate the outcome of a spreading process mostly because they lack the clustering (or modularity) of real-world networks. However we saw how loops, which are the signature of clustering and modularity, affect the onion spectrum of a network. In extreme cases, such as going from a perfect tree to a lattice, loops can change the exponential decay of the spectrum to a linear function. Thus, despite being a tree-like model, the ONE \textit{stretches} the network by considering layers that somehow encode those loops. In other words, while the connection rules of the ONE are strictly local, they mimic the global organization of a given network. For instance, and in contrast to standard rewiring algorithms, the ONE preserves well the diameter of a network, and even offers a good approximation of the entire distribution of shortest path lengths (see Fig.~\ref{fig:paths} and Supplemental Material). Again, the ONE outperforms the CCM while requiring less information.

% =================================================================================================
\section{Conclusion}
% =================================================================================================
%

In summary, we have introduced the onion decomposition that can be useful to (i) characterize complex networks, (ii) identify interesting subgraphs and (iii) approximate their structure---and the dynamics they support---through the onion network ensemble. 

This is done by running an algorithm that scales almost linearly with network size ($O(|E| \log |V|)$) to reveal how quickly a network can be peeled by removing peripheral nodes. The density, or number of nodes, in every layer of the peeling process is called the onion spectrum and characterizes both the core-periphery structure of a network, and how tree-like it is. Moreover, focusing on topological anomalies—layers whose onion spectra go against the general trends—allows us to de novo identify interesting subgraphs within the network. For example, without knowing what precisely we were looking for, we found backbones in networks of websites, governing nodes connecting a set of communities, and dense communities of collaborators within a citation network. This same method could be used to find large communities within an otherwise random network, a sparse periphery to an otherwise dense network, or other interesting structures that have yet to be identified.
%
%Moreover, focusing on topological anomalies---layers whose onion spectra go against the general trends---allows us to identify uniquely interesting subgraphs within the network. For example, we can find large communities within an otherwise random network, or a sparse periphery to an otherwise dense network.

We then showed how all of these properties can be included in a random network ensemble by specifying connection rules that preserve a node's degree and position in the spectrum, and illustrated its potential applications by providing two examples. Future work will focus on solving the properties of that random network ensemble.

%We feel that the two applications investigated---shortest path length distribution, which is a static global property of networks, and SIS prevalence, which is a mostly local dynamical process---illustrate well the range of potential applications for the ONE.

\begin{acknowledgments}
The authors thank Aaron Clauset, Simon DeDeo, and Cristopher Moore for their useful comments. This work has been supported by the Santa Fe Institute, the James S. McDonnell Foundation Postdoctoral Fellowship (LHD), the Santa Fe Institute Omidyar Postdoctoral Fellowship (JAG) and the Fonds de recherche du Qu\'ebec--Nature et technologies (AA).
\end{acknowledgments}

%\newpage \pagestyle{empty}
%\begin{figure}[p]
%\centering
%\includegraphics[page=1,scale=0.9]{LHDetal_OD_2015_Supp.pdf}
%\end{figure}
%\begin{figure}[p]
%\centering
%\includegraphics[page=2,scale=0.9]{LHDetal_OD_2015_Supp.pdf}
%\end{figure}
%\begin{figure}[p]
%\centering
%\includegraphics[page=3,scale=0.9]{LHDetal_OD_2015_Supp.pdf}
%\end{figure}
%\begin{figure}[p]
%\centering
%\includegraphics[page=4,scale=0.9]{LHDetal_OD_2015_Supp.pdf}
%\end{figure}
%\begin{figure}[p]
%\centering
%\includegraphics[page=5,scale=0.9]{LHDetal_OD_2015_Supp.pdf}
%\end{figure}
%\begin{figure}[p]
%\centering
%\includegraphics[page=6,scale=0.9]{LHDetal_OD_2015_Supp.pdf}
%\end{figure}
%\begin{figure}[p]
%\centering
%\includegraphics[page=7,scale=0.9]{LHDetal_OD_2015_Supp.pdf}
%\end{figure}
%\begin{figure}[p]
%\centering
%\includegraphics[page=8,scale=0.9]{LHDetal_OD_2015_Supp.pdf}
%\end{figure}
%\begin{figure}[p]
%\centering
%\includegraphics[page=9,scale=0.9]{LHDetal_OD_2015_Supp.pdf}
%\end{figure}
%\begin{figure}[p]
%\centering
%\includegraphics[page=10,scale=0.9]{LHDetal_OD_2015_Supp.pdf}
%\end{figure}
%\begin{figure}[p]
%\centering
%\includegraphics[page=11,scale=0.9]{LHDetal_OD_2015_Supp.pdf}
%\end{figure}
%\begin{figure}[p]
%\centering
%\includegraphics[page=12,scale=0.9]{LHDetal_OD_2015_Supp.pdf}
%\end{figure}
%\begin{figure}[p]
%\centering
%\includegraphics[page=13,scale=0.9]{LHDetal_OD_2015_Supp.pdf}
%\end{figure}
%\begin{figure}[p]
%\centering
%\includegraphics[page=14,scale=0.9]{LHDetal_OD_2015_Supp.pdf}
%\end{figure}

\end{document}